\documentclass[twocolumn]{aastex63}
\usepackage[normalem]{ulem}
\usepackage{makecell}
\usepackage{graphics}
\usepackage{amsmath}
\usepackage{soul}
\usepackage{CJK}
\usepackage{listings}

\newcommand{\Msun}{\mbox{${\rm M}_{\sun}$}}
\def\CC{{C\nolinebreak[4]\hspace{-.05em}\raisebox{.4ex}{\tiny\bf ++ }}}

\begin{document}
\begin{CJK}{UTF8}{gbsn}
\title{\texttt{wdwarfdate}: A Python Package to Derive Bayesian Ages of White Dwarfs}
\shorttitle{wdwarfdate}
\shortauthors{Kiman et al.}
 
\received{April 21 2022}
\revised{June 6 2022}
\accepted{June 8 2022}
\submitjournal{The Astronomical Journal}

\author[0000-0003-2102-3159]{Rocio Kiman}
\affil{Kavli Institute for Theoretical Physics, University of California, Santa Barbara, CA 93106, USA}
\affil{Department of Physics, Graduate Center, City University of New York, 365 5th Ave, New York, NY 10016, USA}
\affil{Department of Astrophysics, American Museum of Natural History, Central Park West at 79th St, New York, NY 10024, USA}
\affil{Hunter College, City University of New York, 695 Park Ave, New York, NY 10065, USA}
\correspondingauthor{Rocio Kiman}
\email{rociokiman@gmail.com}

\author[0000-0002-8808-4282]{Siyi Xu (许\CJKfamily{bsmi}偲\CJKfamily{gbsn}艺)}
\affil{Gemini Observatory/NSF's NOIRLab, 670 N. A\'ohoku Place, Hilo, HI 96720, USA}

\author[0000-0001-6251-0573]{Jacqueline K. Faherty}
\affil{Department of Astrophysics, American Museum of Natural History, Central Park West at 79th St, New York, NY 10024, USA}

\author[0000-0002-2592-9612]{Jonathan Gagn\'e}
\affiliation{Plan\'etarium Rio Tinto Alcan, Espace pour la Vie, 4801 av. Pierre-de Coubertin, Montr\'eal, Qu\'ebec, Canada}
\affiliation{Institute for Research on Exoplanets, Universit\'e de Montr\'eal, D\'epartement de Physique, C.P.~6128 Succ. Centre-ville, Montr\'eal, QC H3C~3J7, Canada}

\author[0000-0003-4540-5661]{Ruth Angus}
\affil{Department of Astrophysics, American Museum of Natural History, Central Park West at 79th St, New York, NY 10024, USA}
\affil{Center for Computational Astrophysics, Flatiron Institute, 162 5th Avenue, New York, NY 10010 USA}
\affil{Department of Astronomy, Columbia University, 116th St \& Broadway, New York, NY 10027, USA}

\author[0000-0003-2630-8073]{Timothy D. Brandt}
\affil{Department of Physics, University of California, Santa Barbara, Santa Barbara, CA 93106, USA}

\author[0000-0003-2478-0120]{Sarah L. Casewell}
\affil{School of Physics and Astronomy, University of Leicester, University Road, Leicester, LE1 7RH, UK}

\author[0000-0002-1821-0650]{Kelle L. Cruz}
\affil{Department of Physics, Graduate Center, City University of New York, 365 5th Ave, New York, NY 10016, USA}
\affil{Department of Astrophysics, American Museum of Natural History, Central Park West at 79th St, New York, NY 10024, USA}
\affil{Hunter College, City University of New York, 695 Park Ave, New York, NY 10065, USA}
\affil{Center for Computational Astrophysics, Flatiron Institute, 162 5th Avenue, New York, NY 10010 USA}

\begin{abstract} 
White dwarfs have been successfully used as cosmochronometers in the literature, however their reach has been limited in comparison to their potential. We present \texttt{wdwarfdate}, a publicly available Python package to derive the Bayesian age of a white dwarf, based on its effective temperature ($T_{\rm eff}$) and surface gravity ($\log g$).
We make this software easy to use with the goal of transforming the usage of white dwarfs as cosmochronometers into an accessible tool.
The code estimates the mass and cooling age of the white dwarf, as well as the mass and main-sequence age of the progenitor star, allowing for a determination of the total age of the object.
We test the reliability of the method by estimating the parameters of white dwarfs from previous studies, and find agreement with the literature within measurement errors.
By analysing the limitation of the code we find a typical uncertainty of $10\%$ on the total age when both input parameters have uncertainties of $1\%$, and an uncertainty of $25\%$ on the total age when $T_{\rm eff}$ has an uncertainty of $10\%$ and $\log g$ of $1\%$. Furthermore, \texttt{wdwarfdate} assumes single star evolution, and can be applied to calculate the total age of a white dwarf with parameters in the range $1,500  \lesssim T_{\rm eff}  \lesssim 90,000$\,K and $7.9  \lesssim \log g  \lesssim 9.3$. Finally, the code assumes a uniform mixture of C/O in the core and single star evolution, which is reliable in the range of white dwarf masses  $0.45-1.1\,\Msun$ ($7.73  \lesssim \log g  \lesssim 8.8$).

\end{abstract}


\section{Introduction}
\label{sec:intro}

White dwarfs are the perfect candidates to use as cosmochronometers. 
These degenerate objects are formed with high temperatures from the death of stars with masses $<8\,\Msun$, and cool over time.
The cooling process is understood to the point that by deriving the effective temperature ($T_{\rm eff}$ [K]) and surface gravity ($\log g$ [cm ${\rm s^{-2}}$], units for $\log g$ will be implicit from now on to simplify notation) from observational data we can estimate how long it has been cooling with a relatively high accuracy \citep[e.g., ][]{Fontaine2001,Bedard2020}. 
These cooling models, combined with stellar evolutionary models \citep[e.g., ][]{Choi2016,Dotter2016} and the initial-to-final mass relation \citep[IFMR, e.g.,][]{Cummings2018,El-Badry2018,Marigo2020}, provide a reliable method to estimate the total age of a white dwarf.
In addition, with the release of Gaia~DR2 and EDR3 \citep{GaiaCollaboration2016,Collaboration2018,Brown2021}, a large number of new white dwarfs have been discovered --about $10$ times more than previously known-- that have precise proper motions and parallaxes measured by Gaia \citep[e.g., ][]{Fusillo2019,Fusillo2021,McCleery2020}. 

The Gaia mission also allowed for the identification of binaries with unprecedented precision \citep[e.g.,][]{El-Badry2018a,El-Badry2021}.
Given that it can be assumed that the components of a binary system were born at approximately the same time, identifying white dwarfs with wide co-moving companions allows us to calculate the age of the system under the assumption that each object has likely evolved as a single star, without phases of mass transfer or common envelope evolution \citep{Bodenheimer2011}.
Stellar age is one of the most difficult fundamental properties to measure directly; although some methods such as gyrochronology \citep[e.g., ][]{Skumanich1972,Barnes2003,VanSaders2016}, asteroseismology \citep[e.g., ][]{Chaplin2014,Aguirre2017} and isochrone fitting \citep[e.g., ][]{Baraffe2015,Agueros2018,Berger2020} can yield relatively precise ages for single stars, they are limited to specific ranges of masses, or stages of main-sequence evolution, in which they can be applied \citep[e.g., ][]{Chabrier1997,Baraffe2015,Rodriguez2016}. 
For example, none of these methods can be applied to low-mass stars ($<0.6\,\Msun$).
Binary stars provide a powerful tool to estimate stellar ages ``independently of mass": the age of a star of any mass can be estimated if it is co-moving with another object which can be age-dated. 

White dwarf cosmochronology is the subject of extensive literature.
Some examples include \citet{Kalirai2012}, who calculated the ages of white dwarfs in the Milky Way halo to study its formation; \citet{Gagne2018c}, who estimated the age of the ultra-massive white dwarf GD 50 to refine the age of the AB~Dor moving group which GD~50 belongs to; \citet{Anguiano2017}, who estimated white dwarf parameters, including their ages, to study the formation of the Galactic disk; \citet{Catalan2015} who used white dwarf-white dwarf binaries to calibrate the low-mass end of the IFMR; and \citet{Kilic2019}, who used white dwarf ages to study the Milky way inner halo.
In addition to these, other studies (e.g., \citealp{Garces2011} and \citealp{Fouesneau2018}) have estimated the age of main-sequence stars by calculating the age of co-moving white dwarf companions.
Furthermore, previous studies have presented software libraries that can be used to estimate white dwarf ages: \texttt{WD\_models}\footnote{Available at \url{https://github.com/SihaoCheng/WD_models}} (S. Cheng, priv. comm.) is a Python library to estimate white dwarf parameters from Gaia photometry, although it does not provide measurement uncertainties for these parameters; and \texttt{BASE-9}\footnote{Available at \url{https://base-9.readthedocs.io/en/latest/}}  \citep{VonHippel2006,vonHippel2014} is a \CC\ library that estimates star cluster and stellar parameters from photometry, and in particular can be used to estimate white dwarf and progenitor masses and ages.
\texttt{BASE-9} uses a Bayesian technique, combined with a Markov chain Monte Carlo and numerical integration techniques to estimate the posterior probability distributions of the different parameters, including age.
As a consequence, \texttt{BASE-9} results complicated to implement, in particular when only the parameters of the white dwarf are needed. This shows the incredible potential of estimating total ages of white dwarfs, and the lack of tools to make it accessible.

In this paper, we present \texttt{wdwarfdate}, a new Python library to estimate the total age of a white dwarf with the associated uncertainty, from $T_{\rm eff}$ (K), $\log g$, and atmospheric composition. 
The goal of this software is to make the usage of white dwarf as cosmochronometers accessible.
Therefore, this code is designed to be easily used, with warnings for the corresponding limitations of the theoretical models applied in the process. 
The estimation of the white dwarf age is done using Bayesian statistics under the assumptions of a white dwarf core with uniform mixture of C/O and single star evolution. 
\texttt{wdwarfdate} also calculates the mass and cooling age of the white dwarf and the mass and main-sequence age of its progenitor star, along with their respective uncertainties (and their full probability densities) which include scatter in the IFMR and propagate the measurement errors of the white dwarf parameters used as an input.

This paper is divided into the following sections: in Section~\ref{sec:algorithm}, we describe the general method used to estimate the total age of a white dwarf (Section~\ref{subsec:generalidea}), the models that were included for the age determination (i.e., cooling tracks, IFMRs and stellar evolutionary models; Section~\ref{subsec:modelsincluded}), the Bayesian framework on which \texttt{wdwarfdate} relies (Section~\ref{subsec:methods}), and the fast-test mode, an alternative method to estimate the parameters of the white dwarf included in the code (Section~\ref{subsec:fasttestmethod}). 
In Section~\ref{sec:comparativeanalysis}, we apply \texttt{wdwarfdate} to calculate the total ages of a reference sample of age-calibrated white dwarfs to validate our method.
We also estimate the ages of $18$ white dwarfs in wide co-moving pair candidates with M~dwarfs (Section~\ref{subsec:calc_ages}) to provide a sample of age-calibrated M dwarfs for future studies. 
In Section~\ref{sec:constraints}, we detail the constraints and assumptions that are required in using \texttt{wdwarfdate}.  Our results are summarized in Section~\ref{sec:summary}.

\section{Method}
\label{sec:algorithm}

In this section, we describe the sequential steps required to calculate the mass and cooling age of the white dwarf, as well as the mass and the main-sequence age of the progenitor star, and finally the total age of the object. 
\texttt{wdwarfdate} will derive ages using Bayesian statistics --Bayesian age-- by default, but the code also includes a fast-test mode, an alternative method to estimate the parameters of a white dwarf, and we explain both in this section.
We also describe the models in which \texttt{wdwarfdate} relies for the parameter estimation. 
For an example on how to use \texttt{wdwarfdate} see Appendix \ref{sec:run_wdwarfdate}.

\subsection{Outline of the algorithm}
\label{subsec:generalidea}
%

\begin{figure}[ht!]
\begin{center}
\includegraphics[width=0.7\linewidth]{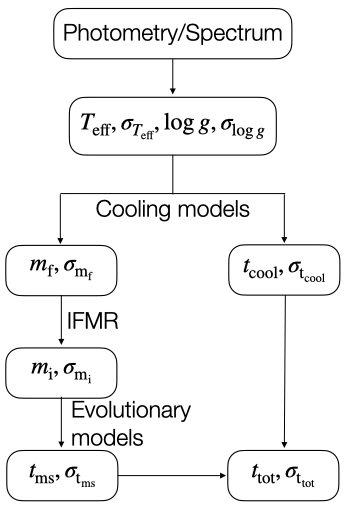}
\caption{Steps required to estimate a white dwarf total age from an effective temperature and surface gravity ($T_{\rm eff}$ and $\log g$). In this diagram, $m_{\rm f}$ and $m_{\rm i}$ are the final mass (mass of the white dwarf) and initial mass (mass of the progenitor star), and $t_{\rm cool}$, $t_{\rm ms}$ and $t_{\rm tot}$ are the cooling age of the white dwarf, the main sequence age of the progenitor star and the total age of the object, respectively. $\sigma _{\rm x}$ indicates the uncertainty of each parameter x.} 
\label{fig:fasttestmethod}
\end{center}
\end{figure}

The calculation of the total age using \texttt{wdwarfdate} relies on a precise measurement of $T_{\rm eff}$ and $\log g$. 
The estimation of these two parameters is done from observational data, and it is described in Section~\ref{subsec:modelsincluded}.
From $T_{\rm eff}$ and $\log g$, the process to estimate the total age of a white dwarf is divided into four stages, summarized in Figure~\ref{fig:fasttestmethod}:
\begin{enumerate}
    \item From the $T_{\rm eff}$ and $\log g$ input values combined with cooling evolutionary tracks for the white dwarf we obtain its mass (final mass, $m_{\rm f}$) and cooling age ($t_{\rm cool}$).
    \item From the $m_{\rm f}$ we estimate the mass of the progenitor star (initial mass, $m_{\rm i}$) using an IFMR, which are, in most cases, empirically calibrated.
    \item With the $m_{\rm i}$ combined with stellar evolution tracks we estimate its evolutionary lifetime, meaning the time since the star formed until it became a white dwarf, which we will call main sequence age or lifetime in this work ($t_{\rm ms}$).
    \item Finally, adding $t_{\rm cool}$ and $t_{\rm ms}$, we obtain the total age of the object ($t_{\rm tot}$).
\end{enumerate}

These steps depend on models which assume single star evolution.
If a white dwarf interacted with another object at any point, then the evolutionary models will not reproduce its parameters correctly \citep[e.g., ][]{Marsh1995,Brown2011}. 
In Section~\ref{sec:constraints} we discuss different scenarios where the single star evolution assumption might not be valid.

\subsection{Models incorporated in \texttt{wdwarfdate}}
\label{subsec:modelsincluded}

The accuracy and precision of the estimated fundamental white dwarf parameters using \texttt{wdwarfdate} depend strongly on how the two input observables were determined, the obtained values, and uncertainties.
In addition, to initialize \texttt{wdwarfdate} the user needs to choose between cooling models for DA white dwarf and non-DA, an IFMR and a stellar evolution model for the progenitor star (metallicity and rotation). 
In this section, we describe the methods generally used to estimate $T_{\rm eff}$ and $\log g$ in the literature and how they affect predictions by \texttt{wdwarfdate}. We also describe the models included in \texttt{wdwarfdate} for each of the choices described above, as summarized in Table~\ref{table:sum_models}.

The determination of temperature and surface gravity generally rely on either a photometric or a spectroscopic approach.
Photometric methods rely on accurate trigonometric parallaxes and high precision photometry \citep[e.g.][]{Koester1979} to reconstruct the spectral energy distribution of a white dwarf, and requires an accurate knowledge of the extinction caused by interstellar dust. Spectroscopic methods, however, rely on atmosphere models to reproduce the detailed profiles of hydrogen or helium lines, and depend on their theoretical line profiles \citep[e.g.][]{Bergeron1992,Beauchamp1999}.
These line profiles in most cases approximate the white dwarf atmosphere as a 1-dimensional object, which can lead to biases in some extreme cases of white dwarf properties, and in most cases ignore the impact of magnetic fields which can be significant for the most massive white dwarfs. 
Recent studies using data from Gaia DR2 and the spectroscopic method show that the results from these two techniques generally agree within the uncertainties \citep{Tremblay2019}, except for $T_{\rm eff} \lesssim 6500$\,K where the parallax-based photometric estimations of $\log g$ were found to be more accurate than spectroscopic estimations \citep{Napiwotzki2020,Kawka2012}.

\begin{figure}[ht!]
\begin{center}
\includegraphics[width=\linewidth]{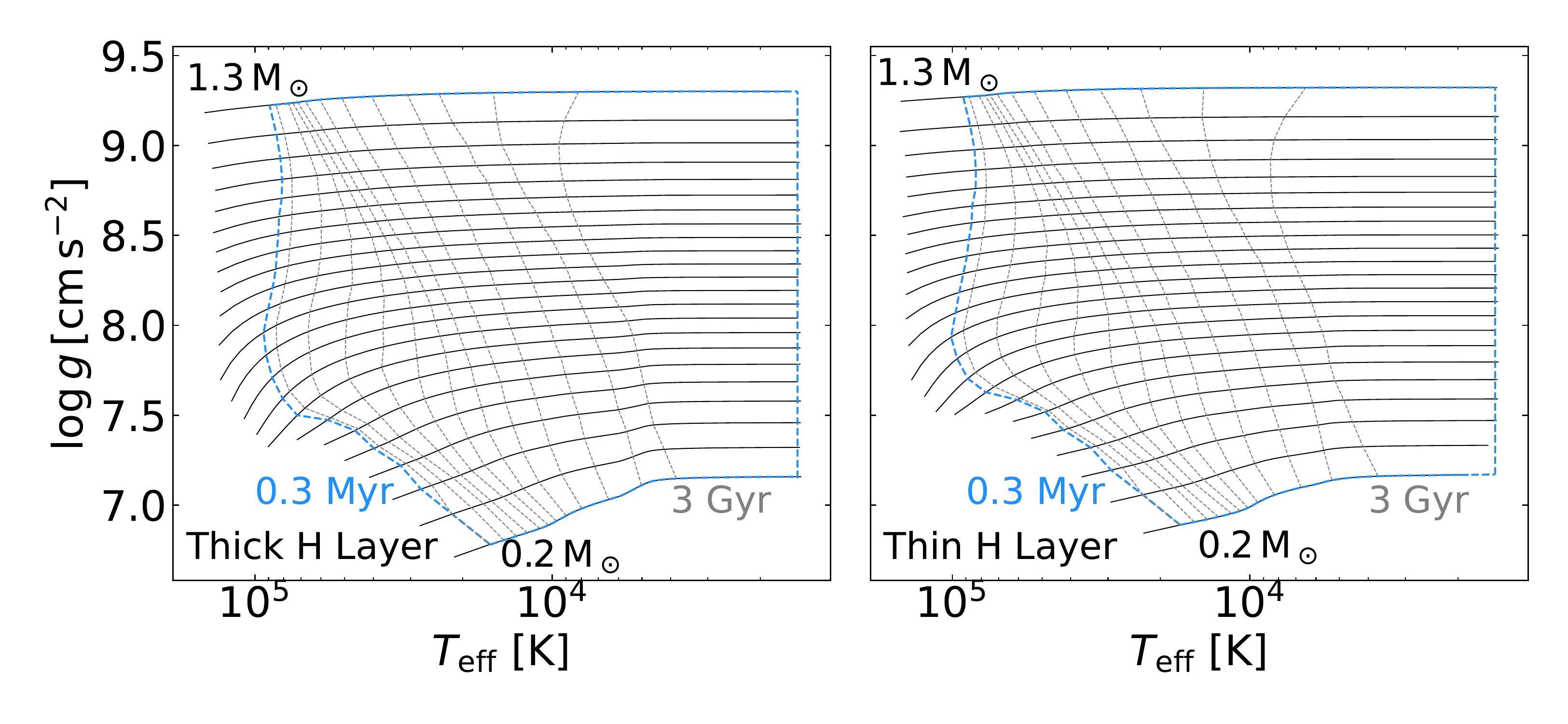}
\caption{White dwarf cooling tracks from \cite{Bedard2020} for thick ($q_{\rm H} \equiv M_{\rm H}/M_{*}=10^{-4}$) and thin ($q_{\rm H} = 10^{-10}$) H layers (left and right panels, respectively). The cooling tracks cover white dwarf masses in the range $0.2-1.3\,\Msun$ in steps of $0.05\,\Msun$. The limits of region in the ($\log g$, $T_{\rm eff}$) plane where \texttt{wdwarfdate} relies on cooling tracks are shown with blue dashed lines. We also included isochrones between $0.3$\,Myr$-3$\,Gyr in gray for comparison and show where the density of isochrones increases.}
\label{fig:cooling_seq}
\end{center}
\end{figure}

\texttt{wdwarfdate} relies on the theoretical evolutionary sequences of the Montreal White Dwarf Group \citep{Bedard2020}.
These models assume that the white dwarf has a core composed of a uniform carbon and oxygen mixture ($X_{\rm C} = X_{\rm O} = 0.5$), surrounded by a He layer (with a fractional mass $q_{\rm He} = M_{\rm He}/M_* = 10^{-2}$) surrounded by an outermost H layer.
The evolutionary tracks assume that this outermost H layer is ``thick" ($q_{\rm H} \equiv M_{\rm H}/M_{*}=10^{-4}$) for hydrogen-atmosphere white dwarfs (DA white dwarf), and ``thin" ($q_{\rm H} = 10^{-10}$) for helium-atmosphere white dwarfs (non-DA white dwarf; \citealp{Bedard2020}).
The publicly available Montreal cooling tracks were calculated for masses in the range $0.2-1.3\,\Msun$ in steps of $0.05\,\Msun$.
These tracks are shown in Figure~\ref{fig:cooling_seq} in black, together with the limits within which \texttt{wdwarfdate} was designed to function in terms of $T_{\rm eff}$ and $\log g$, shown in a blue dashed line.
These limits are based on the following limitations: (1) the lowest temperature covered by the cooling tracks ($1500$\,K); (2) the full range of white dwarf masses covered by the cooling tracks ($0.2-1.3\,\Msun$); and (3) the $0.3$\,Myr isochrone, given that cooling ages below this value are unreliable. 
In summary, the approximate limits within which \texttt{wdwarfdate} will rely on the cooling tracks are: $1,500 \leqslant T_{\rm eff}  \lesssim 100,000$\,K and $7.0  \leqslant \log g  \lesssim 9.3$.
Section~\ref{sec:constraints} details additional limitations of \texttt{wdwarfdate}.

\begin{figure}[ht!]
\begin{center}
\includegraphics[width=\linewidth]{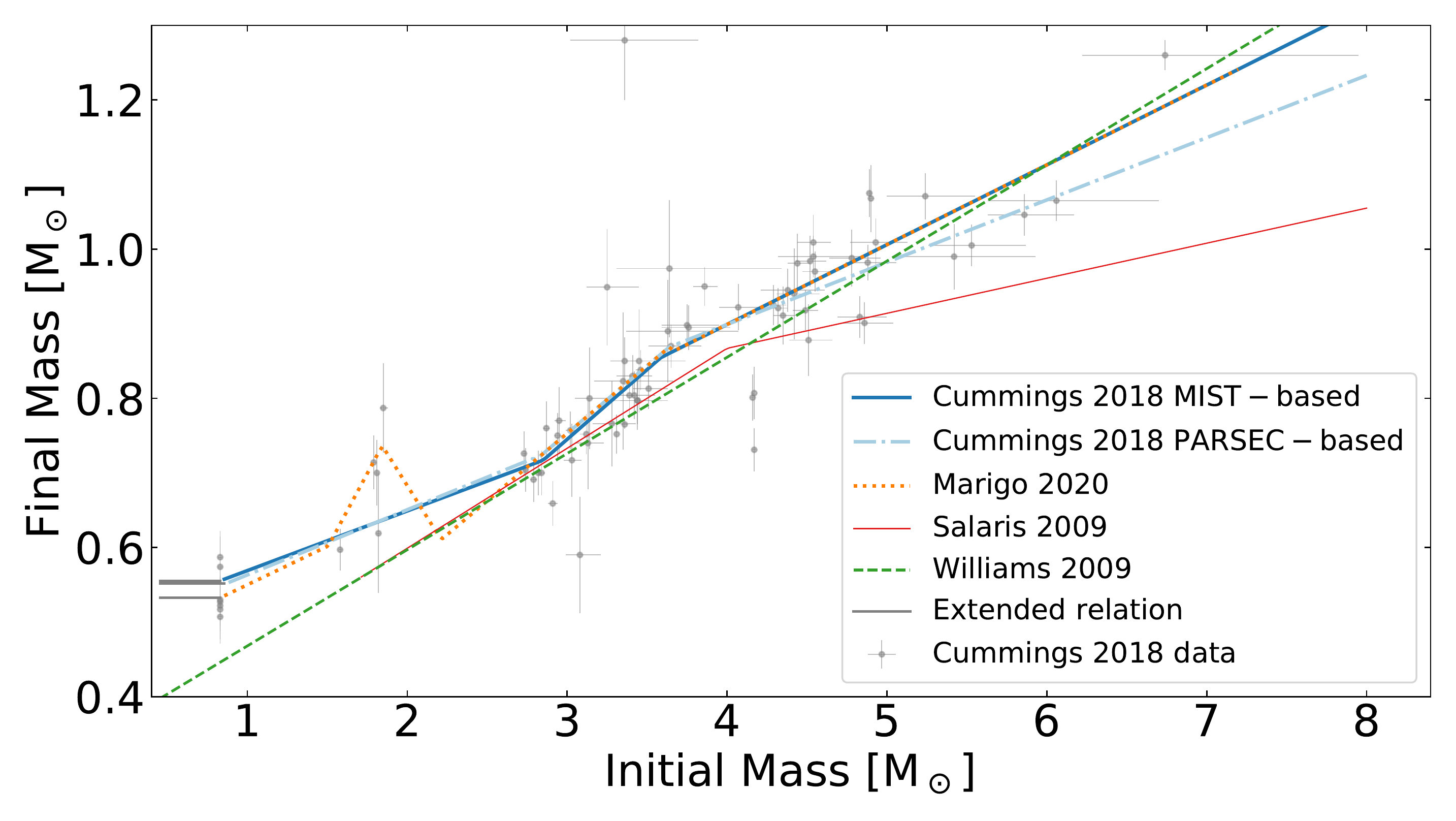}
\caption{Comparison of the initial-to-final mass relations included in \texttt{wdwarfdate}: \citet{Cummings2018} MIST-based (solid blue line) and PARSEC-based (dashed-dotted light blue line), \citet[][dotted orange line]{Marigo2020}, \citet[][solid thin red line]{Salaris2009} and \citet[][dashed green line]{Williams2009}. We include the data from \citet{Cummings2018} as gray points with uncertainties for comparison. As a gray line we show the short extension we included in \texttt{wdwarfdate} of the IFMR from \citet{Cummings2018} and \citet{Marigo2020} at the low-mass end.}
\label{fig:IFMRcomparison}
\end{center}
\end{figure}

To estimate the progenitor mass from the white dwarf mass, or vice versa, the user can choose among the following IFMRs: \citet[MIST- and PARSEC-based]{Cummings2018}, \citet{Marigo2020}, \citet{Salaris2009} and \citet{Williams2009}, shown in Figure~\ref{fig:IFMRcomparison} with semi-empirical data from \citet{Cummings2018}.
Below we describe each of these IFMRs.
\citet{Cummings2018} used white dwarf members of open clusters with known ages to derive the expected progenitor mass and constrain the IFMR of relatively massive (and young) white dwarfs. 
Their analysis is based on a spectroscopic determination of the white dwarfs $T_{\rm eff}$ and $\log g$, which was then used to estimate the cooling age and mass of the white dwarf using the \citet{Fontaine2001} cooling tracks. 
These models are valid for $T_{\rm eff} \lesssim  30,000$\,K, and were updated by the more recent \citet{Bedard2020} cooling tracks.
By subtracting the cooling age from the age of the host open cluster, they obtained the main-sequence age of the progenitor star, which they then used to estimate the progenitor mass using the PAdova and TRieste Stellar Evolution Code \citep[PARSEC,][]{Bressan2012} and the models computed using the Modules for Experiments in Stellar Astrophysics \citep[MESA,][]{Paxton2011,Paxton2013,Paxton2015,Paxton2018}: the MESA Isochrones $\&$ Stellar Tracks \citep[MIST,][]{Choi2016,Dotter2016}.
These two families of stellar evolutionary models resulted in two distinct IFMRs, both of which are included in \texttt{wdwarfdate}.
The difference between the initial masses calculated for the two IFMR are generally within $5\%$. 
By default, \texttt{wdwarfdate} uses the MIST-based IFMR as recommended by \citet{Cummings2018}, due to the non-rotating MIST isochrones estimating more precise ages for young stars ($\sim 100$\,Myr), as tested with the Pleiades cluster. 
However, the PARSEC-based IFMR should be considered especially for high-mass progenitor stars ($>5\,\Msun$), given that the MIST models tend to underestimate masses in this range.

The semi-empirical IFMR from \citet{Marigo2020}, was also obtained from white dwarfs from known clusters. 
In particular, they included seven new white dwarf members of the old open clusters NGC~$752$ ($1.55$\,Gyr) and Ruprecht~$147$ ($2.5$\,Gyr), which allowed them to study the low-mass end (around $0.5\,\Msun$) of the IFMR in detail. 
In addition, \citet{Marigo2020} used a new analysis technique which combines photometric and spectroscopic data to estimate precise $T_{\rm eff}$, $\log g$ and $m_{\rm f}$, and to confirm cluster membership for the white dwarfs.
As shown in Figure~\ref{fig:IFMRcomparison}, they found a kink at low-masses, which they associated with the formation of carbon stars in the Galaxy.
The IFMR from \citet{Salaris2009} also used white dwarfs from known clusters to calibrate the relation, but put special emphasis in the models used, and performed a detailed analysis of the sources of uncertainties and scatter in the relation. 
Finally, the IFMR from \citet{Williams2009} studied white dwarfs from the intermediate-age open cluster M35 (NGC 2168), which provided information to constrain the higher-mass end of the IFMR.

The IFMRs included in \texttt{wdwarfdate} depend on cluster objects that are younger and therefore more massive than average field white dwarfs, which could cause a bias in the relation. 
In addition, as some of the IFMRs were done prior to the Gaia mission, and many of these objects were assumed to be cluster members based on their kinematics and colours, without parallaxes, some of these objects might not be true members, biasing the relation. 
The IFMR will also place an extra boundary in the limits of $T_{\rm eff}$ and $\log g$ for which \texttt{wdwarfdate} can estimate a total age. 
For example, \citet{Cummings2018} MIST-based IFMR was calibrated for white dwarf masses between $0.5-1.3\,\Msun$, which will place a tight constraint specially in the lowest value possible of $\log g$.
We will discuss these limitations further in Section~\ref{sec:constraints}.

The IFMRs from \citet{Cummings2018} and \citet{Marigo2020} are calibrated for $m_{\rm i} > 0.83\,\Msun$, and we included them in the code extending this limit to $0.45\,\Msun$, assuming a constant $m_{\rm f}$ between $0.45< m_{\rm i}<0.83$ (gray line in Figure~\ref{fig:IFMRcomparison}). 
This extension allows \texttt{wdwarfdate} to estimate parameters for white dwarfs with masses $m_{\rm f} \lesssim 0.6\,\Msun$, which is the peak of the white dwarf mass distribution \citep[e.g., ][]{Genest-Beaulieu2019}.

To estimate the main sequence age from the initial mass, or progenitor mass, we adopted in \texttt{wdwarfdate} the MIST isochrones.
These tracks cover a mass range between $0.1-300\,\Msun$ with a step of $0.05\,\Msun$, and an age range from $0.1$\,Myr to $13.8$\,Gyr.
To select the optimal stellar evolution track, the user has to choose the metallicity and rotation of the progenitor star among the following options: ${\rm [Fe/H]}=\{-4,-1,0,0.5\}$ (dex), $[{\rm \alpha/Fe}]=0$, and ${\rm v/vcrit} = \{0.0,0.4\}$. 
We will discuss the effects of using different tracks on the total age in Section~\ref{sec:constraints}.

\begin{deluxetable}{ll}[ht!]
\tabletypesize{\scriptsize}
\tablecaption{Models included in \texttt{wdwarfdate}. \label{table:sum_models}}
\tablehead{\colhead{Models} & \colhead{Model included}    
}\startdata 
Cooling tracks & \citet{Bedard2020}\tablenotemark{a}: \\
 & DA and non-DA white dwarfs\\
IFMR & \citet{Marigo2020} \\
 & \citet{Cummings2018} MIST  \\
 & \citet{Cummings2018} PARSEC  \\
 & \citet{Salaris2009} \\ 
 & \citet{Williams2009} \\
Stellar evolution tracks & MIST isochrones\tablenotemark{b} \\  
& \citep{Choi2016,Dotter2016}: \\
 & $[{\rm Fe/H}]=\{-4,-1,0,0.5\}$ \\
 & $[{\rm \alpha/Fe}]=0$ \\
 & ${\rm v/vcrit} = \{0.0,0.4\}$\\
\enddata 
\tablenotetext{a}{Available online \url{http://www.astro.umontreal.ca/~bergeron/CoolingModels/}}
\tablenotetext{b}{Available online \url{http://waps.cfa.harvard.edu/MIST/}}
\end{deluxetable}

\subsection{Bayesian framework}
\label{subsec:methods}

\begin{figure}[ht!]
\begin{center}
\includegraphics[width=.7\linewidth]{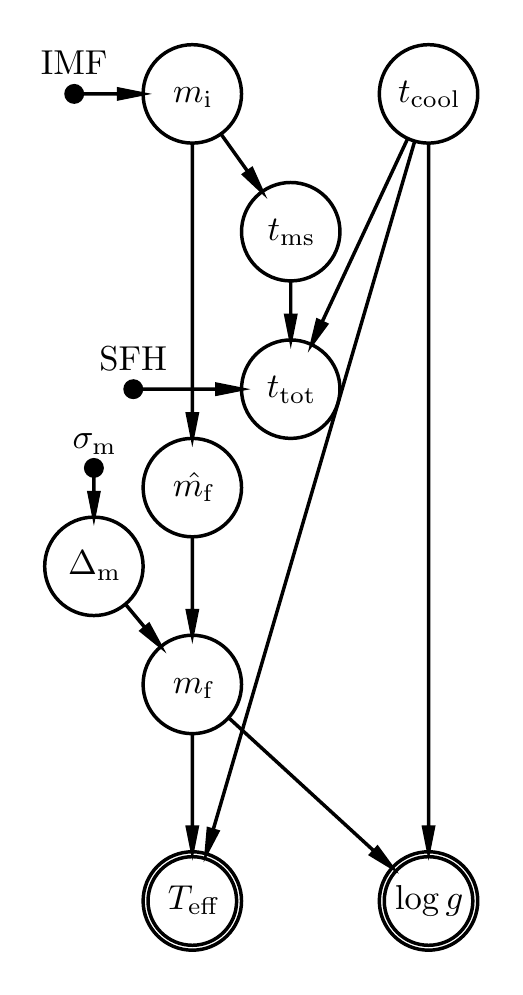}
\caption{Probabilistic Graphical Model for the Bayesian method of \texttt{wdwarfdate}. In short, $T_{\rm eff}$ and $\log g$ are calculated from the parameters being sampled ($m_{\rm i}$, $t_{\rm cool}$ and $\Delta _{\rm m}$), and these values are compared to the input parameters. In the figure, $\hat{m_{\rm f}}$ is the scattered final mass (mass of the white dwarf) calculated from the initial mass using an IFMR, and $\Delta _{\rm m}$ is an extra parameter to take into account the scatter in this relation. Therefore, $m_{\rm f}$ is the true final mass, calculated by adding $\Delta _{\rm m}$ to $\hat{m_{\rm f}}$.} 
\label{fig:pgm}
\end{center}
\end{figure}

In this section, we describe the Bayesian method implemented in \texttt{wdwarfdate}. 
Our code derives Bayesian ages of white dwarfs, with a posterior probability given by 

\begin{equation}
    p(m_{\rm i},\log _{10}t_{\rm cool},\Delta _{\rm m}|T_{\rm eff},\log g) \label{eq:posterior_short}
\end{equation}

\noindent where $m_{\rm i}$ is the initial mass, or the mass of the progenitor star, and $t_{\rm cool}$ is the cooling age of the white dwarf. 
$\Delta _{\rm m}$ is an extra parameter which takes into account the scatter in the IFMR of white dwarfs, and it is described below. 
$T_{\rm eff}$ and $\log g$ are the inputs for the effective temperature and surface gravity, respectively. 
To simplify notation we did not include uncertainties in Equation~\ref{eq:posterior_short}, although these are also required in the code.
Using Bayes theorem, the posterior probability in Equation \ref{eq:posterior_short} can be written as

\begin{equation}
\begin{split}
    p(m_{\rm i},\log _{10}t_{\rm cool},&\Delta _{\rm m}|T_{\rm eff},\log g) \propto \\
    & p(T_{\rm eff},\log g|m_{\rm i},\log _{10}t_{\rm cool},\Delta _{\rm m}) \\
    & p(\log _{10}t_{\rm cool})p(m_{\rm i})p(\Delta _{\rm m}),  \label{eq:posterior}
\end{split}
\end{equation}

\noindent where $p(\log _{10}t_{\rm cool})$, $p(m_{\rm i})$ and $p(\Delta _{\rm m})$ are the priors on the cooling age, the initial mass and $\Delta _{\rm m}$ respectively, which are described below, and 

\begin{equation}
\begin{split}
p(&T_{\rm eff},\log g|m_{\rm i},\log _{10}t_{\rm cool},\Delta _{\rm m}) \propto \\
&\exp \left[ -\frac{(\overline{T_{\rm eff}}-T_{\rm eff})^2}{2\sigma_{T_{\rm eff}}^2} \right] 
\exp \left[ -\frac{(\overline{\log g}-\log g)^2}{2\sigma_{\log g}^2} \right] 
\label{eq:likelihood}
\end{split}
\end{equation}
    
\noindent is the likelihood.
$\overline{T_{\rm eff}}$ and $\overline{\log g}$ in Equation~\ref{eq:likelihood} are the modeled effective temperature and surface gravity of the white dwarf. 
These modeled quantities are calculated from the three parameters that are being sampled ($m_{\rm i}$, $t_{\rm cool}$ and $\Delta _{\rm m}$) as shown by the probabilistic graphical model in Figure~\ref{fig:pgm}, which represents the posterior probability of the problem.
In summary: 1) From the initial mass ($m_{\rm i}$) we estimate a main sequence age ($t_{\rm ms}$); 2) by adding $t_{\rm ms}$ to the cooling age ($t_{\rm cool}$) we obtain the total age ($t_{\rm tot}$) of the white dwarf; 3) from $m_{\rm i}$ and the delta parameter ($\Delta _{\rm m}$) we estimate a final mass ($m_{\rm f}$, described below); 4) from $m_{\rm f}$ and $t_{\rm cool}$ we estimate the modeled effective temperature and surface gravity ($\overline{T_{\rm eff}}$ and $\overline{\log g}$) which are compared to the input parameters ($T_{\rm eff}$ and $\log g$) to estimate the posterior probability. 
The models used in each step are described in Section~\ref{subsec:modelsincluded}.

The $\Delta _{\rm m}$ parameter in Equations~\ref{eq:posterior_short}, \ref{eq:posterior} and \ref{eq:likelihood} was included to model the scatter in the IFMR.
This scatter has different causes including contamination from non-cluster members, measurement errors, metallicity variations, inconsistencies with the models used and age determination of the clusters \citep[e.g.,][]{Weidemann2000,Williams2009,Casewell2009}.
The parameter $\Delta _{\rm m}$ is drawn from a normal distribution centered at zero with standard deviation $\sigma _{\rm m}$, so that

\begin{equation}
    p(\Delta _{\rm m}) = \mathcal{N}(0,\sigma _{\rm m}). \label{eq:delta_m}
\end{equation}

\noindent We estimated the parameter $\sigma _{\rm m}$ from the \citet{Cummings2018} data, and adopted $\sigma _{\rm m} = 0.03\,\Msun$.
The inclusion of the scatter in the posterior is indicated in Figure~\ref{fig:pgm} as $\hat{m}_{\rm f} = m_{\rm f}+\Delta _{\rm m}$, where $m_{\rm f}$ is the true mass of the white dwarf and $\hat{m}_{\rm f}$ is the scattered mass, because it was calculated using the IFMR.

The priors indicated in the posterior probability of Equation~\ref{eq:posterior} are given by the Star Formation History (SFH) for $t_{\rm cool}$ and the Initial Mass Function (IMF) for $m_{\rm i}$.
We assume a constant SFH \citep[e.g., ][]{Madau2014} and applied it as a flat prior on $t_{\rm tot}$, where we required $t_{\rm tot}<15$\,Gyr to avoid a sharp cut at the age of the Universe in the age probability distributions. 
As $t_{\rm tot} = t_{\rm cool} + t_{\rm ms}$, the Jacobian of the coordinate transformation is $1$, and this prior could be used without extra modifications. 
In addition, we assumed the IMF to be $m_{\rm i}^{-2.3}$ \citep{Kroupa2001,Chabrier2003}, which describes the range of stellar masses expected to form a white dwarf.

\texttt{wdwarfdate} obtains the probability distributions by explicitly integrating over the likelihood and priors to perform the normalization.
We set the code to evaluate the posterior in a grid made of discrete arrays for each parameter: $m_{\rm i}$, $\log _{10} t_{\rm cool}$ and $\Delta _{\rm m}$.
The code first performs a wide grid evaluation with the parameters ranging between $0.3\,{\rm Myr} < t_{\rm cool} < 15\,{\rm Gyr}$, $0.1\,\Msun<m_{\rm i}<10\,\Msun$ and $-0.1\,\Msun<\Delta _{\rm m}<0.1\,\Msun$. 
Then it obtains the probability distribution for each parameter by setting the minimum and maximum evaluation limits according to where the probability is higher. 
These limits can also be adjusted by the user. 
We use these results combined with the models described in Section~\ref{subsec:modelsincluded} to derive probability distributions for the rest of the parameters (total age, final mass and main sequence age).

\subsection{Fast-test method}
\label{subsec:fasttestmethod}

One of the limitations of \texttt{wdwarfdate} is given by the IFMR: the total age of a white dwarf can be estimated only in the cases where its mass is between the allowed limits of the IFMR.
For the cases where a total age cannot be estimated but the $T_{\rm eff}$ and $\log g$ are within the limits of the cooling tracks, we included in \texttt{wdwarfdate} a \textit{fast-test} method to estimate the final mass and cooling age only.
For this method, \texttt{wdwarfdate} generates two normal distributions for the effective temperature and surface gravity, using the input values ($T_{\rm eff}$ and $\log g$) as means, and the uncertainties ($\sigma _{T_{\rm  eff}}$ and $\sigma _{\log g}$) as the standard deviations of the distributions such as

\begin{equation}
\begin{split}
    X_{T_{\rm eff}} &\sim \mathcal{N}(T_{\rm eff},\sigma _{T_{\rm  eff}})\\
    X_{\log g} &\sim \mathcal{N}(\log g,\sigma _{\log g}).
\end{split}
\end{equation}

\noindent The code does a Monte Carlo uncertainty propagation by running the two distributions through the process described in Section~\ref{subsec:generalidea} and in Figure~\ref{fig:fasttestmethod}, and obtains a distribution for each of the parameters of interest: final mass, initial mass, cooling age, main sequence age and total age. 
\texttt{wdwarfdate} will automatically run this method when the final mass of the white dwarf is outside of the allowed ranges by the IFMR. 
The main difference between the fast-test and the Bayesian methods are the priors: the fast-test method does not include the IMF and SFH as priors on the initial mass and the total age, respectively, but it does constrain the total age to be smaller than $15$\,Gyr.

\begin{figure}[ht!]
\begin{center}
\includegraphics[width=\linewidth]{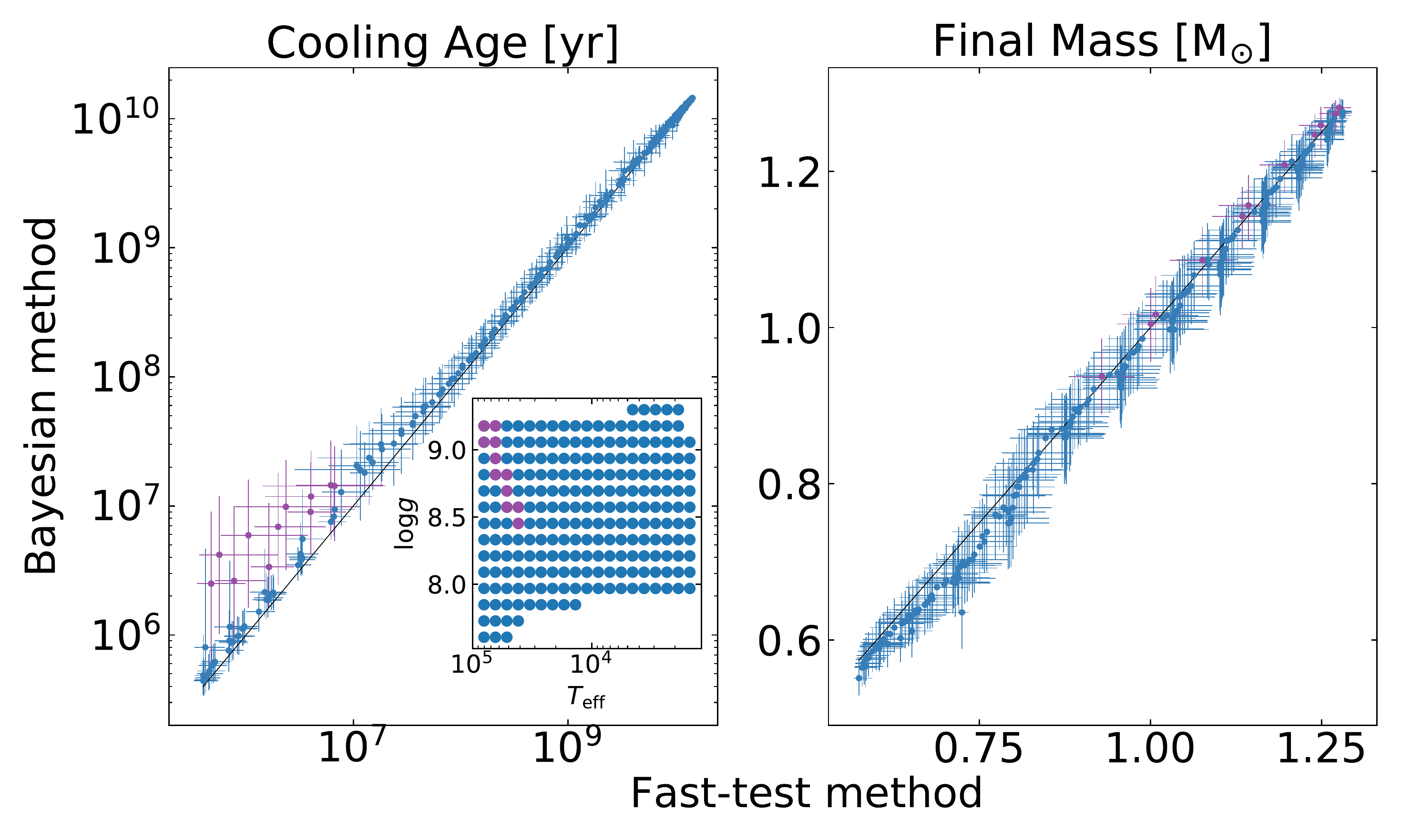}
\caption{Comparison of the fast-test and Bayesian methods included in \texttt{wdwarfdate} for the cooling age and final mass (left and right panels, respectively). We included the one-to-one relation in a black solid line. We run a grid of white dwarfs with $T_{\rm eff}=1,500-100,000$\,K and $\log g=7-9.3$ and uncertainties of $10\%$ and $1\%$, respectively. This comparison shows that the estimation of final mass and cooling age using both methods is similar. The purple points in both panels represent calculations of cooling age that differ more than $1\sigma$ between the two methods. The small panel of $T_{\rm eff}$ and $\log g$ shows that these points are situated in a dense area of isochrones (Figure~\ref{fig:cooling_seq}), which makes the likelihood less constrained. As a consequence, the prior has more influence and makes the Bayesian cooling ages larger.} 
\label{fig:fasttestbayesian}
\end{center}
\end{figure}

To test that the final mass and cooling age estimated with the fast-test and the Bayesian methods are comparable, we run a grid of white dwarfs with $T_{\rm eff}=1,500-100,000$\,K and $\log g=7-9.3$ and uncertainties of $10\%$ and $1\%$ respectively, using the two methods.
For both methods we used the DA white dwarf cooling sequences, the \citet{Cummings2018} MIST-based IFMR, the MIST tracks with ${\rm [Fe/H]}=0$ and ${\rm v/vcrit}=0.0$, and the uncertainties were calculated using the $16^{\rm th}$ and $84^{\rm th}$ percentile of the distributions.
The results show that the final mass estimations agree within the uncertainties between both methods (right panel Figure~\ref{fig:fasttestbayesian}), and the cooling age estimations agree in most cases (left panel Figure~\ref{fig:fasttestbayesian}).
We color-coded in purple the points for which the cooling age differs by more than $1\sigma$, and we found that these points agree in the final mass within the uncertainty, although the cooling age values seem to be more discrepant as the cooling age decreases. 
This discrepancy does not depend on the cooling age because there are some white dwarfs with small cooling ages which have the same result in the fast-test and Bayesian method. 
The purple points on Figure~\ref{fig:fasttestbayesian} have some of the highest temperatures and surface gravities, as shown by the $(T_{\rm eff}-\log g)$ inset diagram in the left panel of Figure~\ref{fig:fasttestbayesian}. 
Moreover, the purple points follow the area where the cooling tracks are closer to each other in Figure~\ref{fig:cooling_seq}.
This increases the uncertainty in the cooling age, making the likelihood probability distribution function (PDF) of these objects wider.
Therefore, the prior, in particular the IMF, has a bigger effect on the estimated values, making the cooling ages estimated with the Bayesian method larger, given that it favors smaller initial masses.

The final masses between $0.65-0.8\,\Msun$ in the right panel of Figure~\ref{fig:fasttestbayesian} are slightly lower with the Bayesian method than with the fast-test method.
This effect is also caused by the prior, in a similar way as described above.
By inspecting the cooling tracks in Figure~\ref{fig:cooling_seq}, we noticed that the cooling tracks in the mass range $0.65-0.8\,\Msun$ are closer to each other, making the likelihood less constrained.
The differences for the final mass and cooling age described above show the advantage of applying the Bayesian method which adds extra information using the priors (IMF and SFH), unlike the fast-test method which provides the highest likelihood value.

In summary, when the uncertainties in $T_{\rm eff}$ and $\log g$ are small ($\sim 10\%$ and $\sim 1\%$, respectively) the fast-test and the Bayesian method are comparable in the estimation of final masses and cooling ages, except for white dwarfs with both $T_{\rm eff} > 40,000$\,K and $\log g>8.4$, which coincide with dense track areas. However, these exceptions represent only $2\%$ of our sample, and some have extremely high temperatures, therefore these cases are not common.

\section{Results of applying \texttt{wdwarfdate} to literature and new white dwarfs}
\label{sec:comparativeanalysis}

In this section we compare the results from \texttt{wdwarfdate} with those from previous studies.
The goal of this section is to test the validity of \texttt{wdwarfdate} results.
We also calculate the ages of white dwarfs in candidate binaries with M~dwarfs which make a new set of age calibrators.
Unless indicated otherwise, the estimations were made using the Bayesian method and the \citet{Cummings2018} MIST-based IFMR, with the DA white dwarf cooling sequences, and ${\rm [Fe/H]}=0$ and ${\rm v/vcrit}=0.0$ stellar evolution models. The uncertainties were calculated using the $16^{\rm th}$ and $84^{\rm th}$ percentile of the distributions.

\subsection{Comparison to white dwarfs from clusters from \citet{Cummings2018}}
\label{subsec:comparisonya}

\citet{Cummings2018} used a sample of $79$ white dwarfs from $13$ different clusters of known ages with $T_{\rm eff}$ and $\log g$ measurements from the literature to calibrate the IFMR. 
In addition, the authors re-measured $T_{\rm eff}$ and $\log g$ for $8$ of the white dwarfs. 
\citet{Cummings2018} is one of the most comprehensive studies to-date of white dwarfs from clusters.
They used the cooling tracks from the Montreal White Dwarf Group to estimate final masses and cooling ages from $T_{\rm eff}$ and $\log g$, and the MIST isochrones to estimate initial masses from the cooling and total ages of the objects (See Section~\ref{subsec:modelsincluded} for a detailed description of their work). 
Given that they estimated all the parameters which can be obtained with our code using a similar procedure, \citet{Cummings2018} is an ideal study to compare the performance of \texttt{wdwarfdate}.

\begin{figure}[ht!]
\begin{center}
\includegraphics[width=\linewidth]{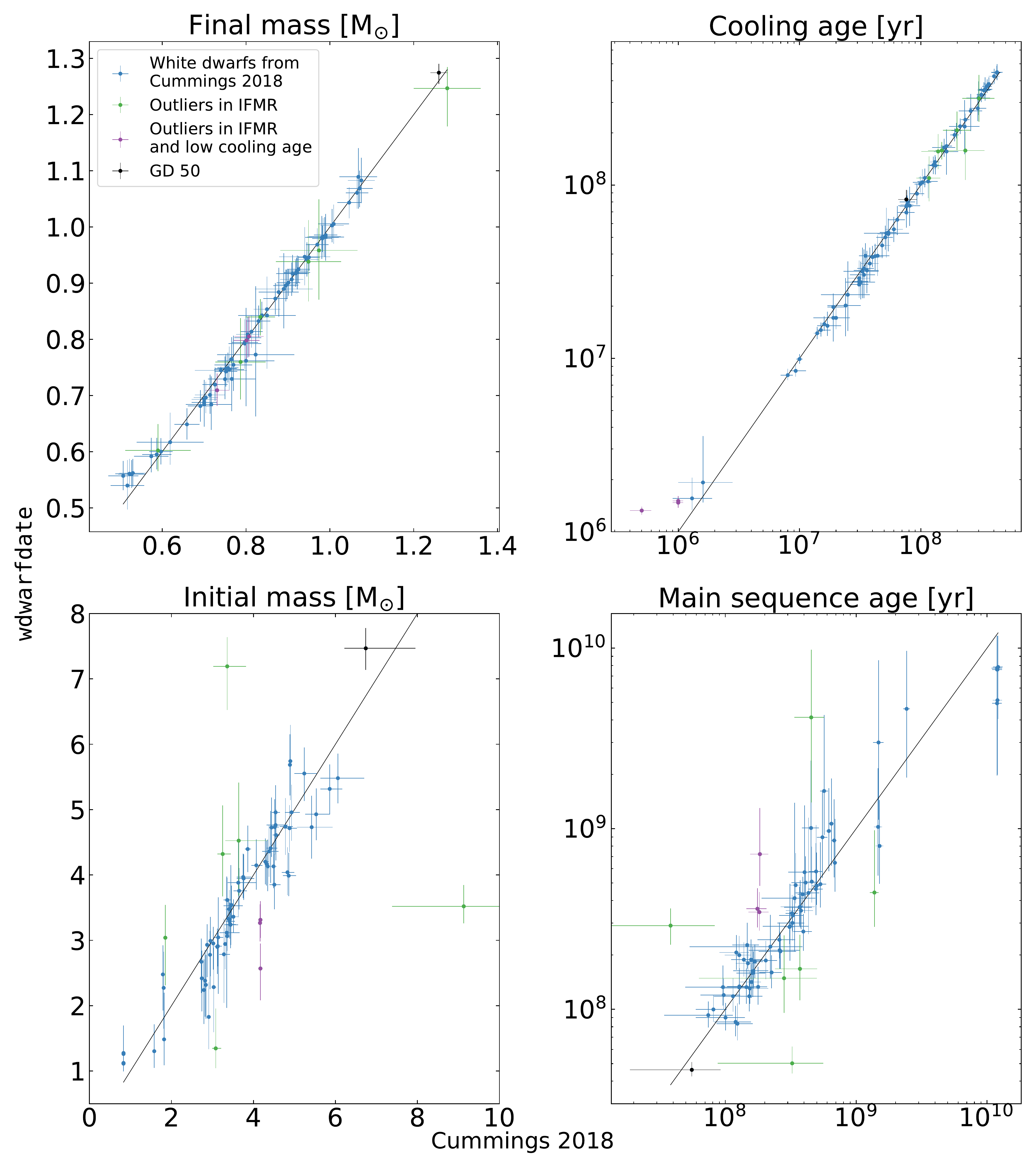}
\caption{We run the white dwarfs in \citet{Cummings2018} using \texttt{wdwarfdate} to compare the results. Most of the parameters are within $1\sigma$ and we color-coded the rest. Both the green and purple points are outliers in the IFMR. In addition, the three purple white dwarfs with higher cooling age have high effective temperatures, which are better modeled with the new cooling tracks implemented in our code \citep{Bedard2020}. The black dot is the massive white dwarf GD 50.}
\label{fig:comparison_cummings_parameters}
\end{center}
\end{figure}

We run \texttt{wdwarfdate} on all the stars in \citet{Cummings2018} using the $T_{\rm eff}$ and $\log g$ and uncertainties published in their work. 
We compared final mass, cooling age, initial mass and main sequence age in Figure~\ref{fig:comparison_cummings_parameters}.
Most of the values agree within $1\sigma$ with their results.
The uncertainties calculated with \texttt{wdwarfdate} are comparable to the estimated by \citet{Cummings2018} for the cooling age and final mass. 
For the main sequence age and initial mass, the uncertainties are smaller in the \citet{Cummings2018} results because they were propagated from the age of the cluster which has a small uncertainty, and not the IFMR as in \texttt{wdwarfdate}. 

We color-coded the white dwarfs for which any of the parameters in Figure~\ref{fig:comparison_cummings_parameters} did not agree within $1\sigma$ to study them in detail. 
The black dot is the white dwarf GD 50, and although the parameters agree within the uncertainties it is color-coded because it is discussed below.
The white dwarfs color-coded in green and purple are outliers in the IFMR (Figure~\ref{fig:IFMRcomparison}), which explains why the initial mass and main sequence age do not agree. 
We refer to \citet{Cummings2018} for the analysis of the outliers.
The final mass and cooling age for the green objects agree within uncertainties, which is expected because we are using the same $T_{\rm eff}$ and $\log g$ to estimate them.
In the case of the three white dwarfs color-coded in purple, \texttt{wdwarfdate} estimated a higher cooling age than expected according to the parameters in \citet{Cummings2018}.
However, these objects have $T_{\rm eff}>50,000$\,K, and the cooling tracks used in \citet{Cummings2018} \citep{Fontaine2001} are not suitable for temperatures $>30,000$\,K. 
We confirmed our calculations of the cooling ages with the Montreal White Dwarfs Data Base \citep[MWDD, ][]{2017ASPC..509....3D}\footnote{\url{https://www.montrealwhitedwarfdatabase.org/}}, which contains the updated models by \citet{Bedard2020}, which we implemented in \texttt{wdwarfdate}. 

To compare the total age estimated with \texttt{wdwarfdate} with the cluster age, we excluded the outliers (points color-coded in green and purple) in Figure~\ref{fig:comparison_cummings_parameters}. 
The total estimated age agrees within $1\sigma$ with the cluster age for $65\%$ of the white dwarfs, and within $2\sigma$ for $94\%$. 
We also calculated a median relative difference of $12\%$ between the median age for each white dwarf that was not an outlier and the age according to the cluster membership, which shows \texttt{wdwarfdate} is doing a good job estimating these total ages. 


\begin{figure}[ht!]
\begin{center}
\includegraphics[width=\linewidth]{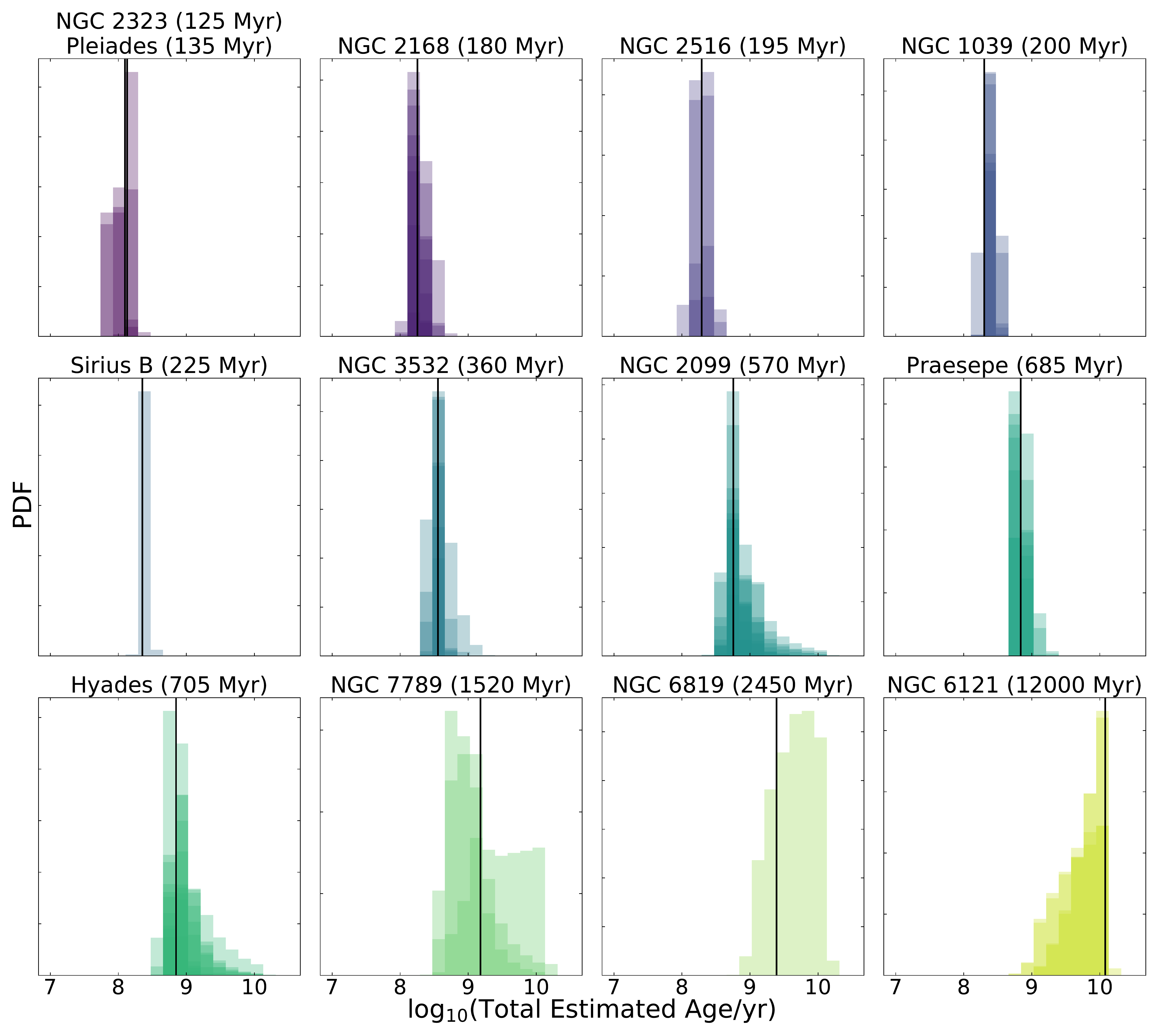}
\caption{We compare the total age posterior PDF obtained with \texttt{wdwarfdate} for the white dwarfs in \citet{Cummings2018}, with the age of the cluster they belong to. Also, we divided the white dwarfs according to their membership and added in a black vertical line the age of the group. We found good agreement between the age of the cluster and the age derived with \texttt{wdwarfdate}.}
\label{fig:comparison_cummings_total_age}
\end{center}
\end{figure} 

The white dwarf color-coded in black is GD $50$, which is a young and ultramassive white dwarf of $1.28 \pm 0.08\,\Msun$. 
For this object we obtained a main sequence age of $46.21 ^{+4.76}_{-3.74}$\,Myr, a cooling age of $82.62^{+11.19}_{-14.08}$\,Myr, a total age of $128.71^{+10.54}_{-11.92}$\,Myr, a final mass of $1.27\pm0.02\,\Msun$, and an initial mass of $7.5\pm0.3\,\Msun$.
Our progenitor mass calculation differs slightly from the value in \citet{Cummings2018}, but agrees with the results of \citet{Gagne2018c}, that calculated a mass of $7.8\pm0.6\,\Msun$. 
They found GD $50$ is likely member of the AB Doradus moving group, and performed a detailed estimation of its parameters using the Montreal white dwarf evolutionary models and the MIST models, accounting for all possible C/O/Ne core compositions, and finding a total age of $117\pm26$\,Myr. 
Our result is between this value and the one estimated by \citet{Cummings2018} that assumed GD~$50$ belongs to the the Pleiades (based on \citealt{2006MNRAS.373L..45D}) and assigned it an age of $135\pm35$\,Myr, which is slightly larger than the current measured age of this cluster ($112 \pm 5$\,Myr; \citealt{Dahm2015}). 
We also used \texttt{wdwarfdate} on GD~$50$ using the PARSEC-based IFMR from \citet{Cummings2018}, as was suggested in Section~\ref{subsec:modelsincluded} for progenitor masses $>5\,\Msun$, and we obtained a total age of $114.61^{+12.72}_{-11.68}$\,Myr, significantly closer to the results of \citet{Gagne2018c}.

\subsection{Comparison to white dwarfs from M$67$ from \citet{Canton2021}}
\label{subsec:comparison_low_mass}

\begin{figure}[ht!]
\begin{center}
\includegraphics[width=\linewidth]{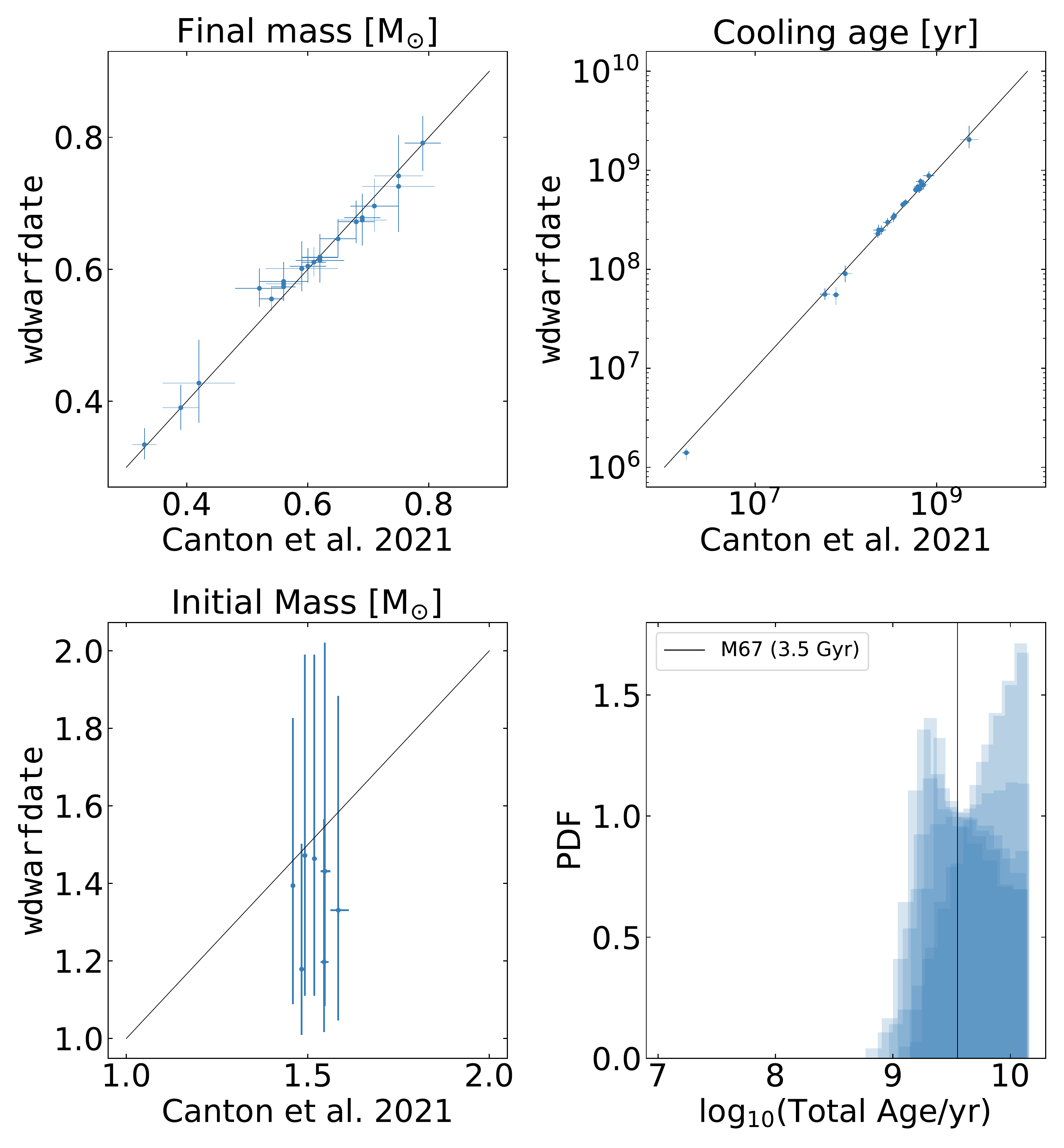}
\caption{Comparison of the performance of \texttt{wdwarfdate} at lower-masses, with the results from \citet{Canton2021}. These white dwarfs belong to the M$67$ cluster ($3.5$\,Gyr). For the final mass, cooling age and initial mass we show the one-to-one line in black, and for the total age we show the age distributions with the age of the cluster in a black line. For the total age and initial mass we only show the vetted white dwarfs from \citet{Canton2021}. We found good agreement between our results and their study. The uncertainties in the initial mass calculated by \citet{Canton2021} are smaller than the point in some cases because they used a well constrained total age to estimate these masses.}
\label{fig:comparison_canton_parameters}
\end{center}
\end{figure}

To test the accuracy of \texttt{wdwarfdate} at lower masses, we estimated the parameters of the white dwarfs from \citet{Canton2021}.
The authors measured $T_{\rm eff}$ and $\log g$ from high resolution spectra for $22$ white dwarfs from the M$67$ cluster which is $3.5$\,Gyr and solar metallicity. 
They used the cooling tracks from \citet{Fontaine2001} to estimate final masses from these parameters, and the PARSEC isochrones to estimate initial masses from the age of the cluster.
They estimated initial masses only for a smaller vetted sample, and discarded the rest because, for example, they were potential He-core white dwarfs or potential blue straggler remnants.
Using their measured $T_{\rm eff}$ and $\log g$, and uncertainties with \texttt{wdwarfdate}, we found good agreement for final mass and cooling age between our results and \citet{Canton2021} --as shown in Figure \ref{fig:comparison_canton_parameters}-- for all but two white dwarfs. 
However, our results for these two white dwarfs agree with the MWDD, therefore the differences are likely due to \citet{Canton2021} using an older version of the Montreal cooling tracks \citep{Fontaine2001}.
In addition, we compared our estimation of the total age and the initial mass for the vetted sample from \citet{Canton2021}, and found that our results for both parameters agree within $1\sigma$, as shown by the bottom panels in Figure~\ref{fig:comparison_canton_parameters}.
The uncertainties on the total age of the white dwarfs are large ($\pm 5$\,Gyr) which is expected given that these are low-mass white dwarfs and the largest contribution to the total age comes from the main sequence age, which tends to have higher uncertainties. 
Furthermore, the uncertainties on the initial mass calculated by \citet{Canton2021} are smaller because they used the known total age to estimate these masses, which is well constrained.
To further check our code, we run \texttt{wdwarfdate} using the IFMR from \citet{Marigo2020} which differs from \citet{Cummings2018} for low-masses, and found the same results for the median value of the parameters, although the shape of the posterior was different in some cases, due to the change in the shape of the IFMR.

\subsection{Wide, co-moving white dwarf companions}
\label{subsec:calc_ages}

We used \texttt{wdwarfdate} to estimate ages of $30$ white dwarfs identified to be candidate wide co-movers with M~dwarfs by \citet{Kiman2021}. 
These M~dwarfs constitute a new set of age-calibrators and we refer the reader to  \citet{Kiman2021} for information about the main-sequence co-moving star.
\citet{Kiman2021} estimated the white dwarf ages using \texttt{wdwarfdate}, and the $T_{\rm eff}$ and $\log g$ calculated by \citet{Fusillo2019} with photometry from Gaia DR2. 
For our calculations we used the updated version of that work \citep{Fusillo2021}, that took advantage of the precise photometry and parallaxes from Gaia EDR3 and estimated $T_{\rm eff}$ and $\log g$ for $359,073$ high-confidence white dwarf candidates.
However, we did not find improvement between Gaia DR2 and EDR3 in the relative uncertainty of $T_{\rm eff}$ and $\log g$ for the $30$ white dwarfs, and we found a relative difference for the values of $8.4\%$ and $2.2\%$ respectively. 

\begin{figure}[ht!]
\begin{center}
\includegraphics[width=\linewidth]{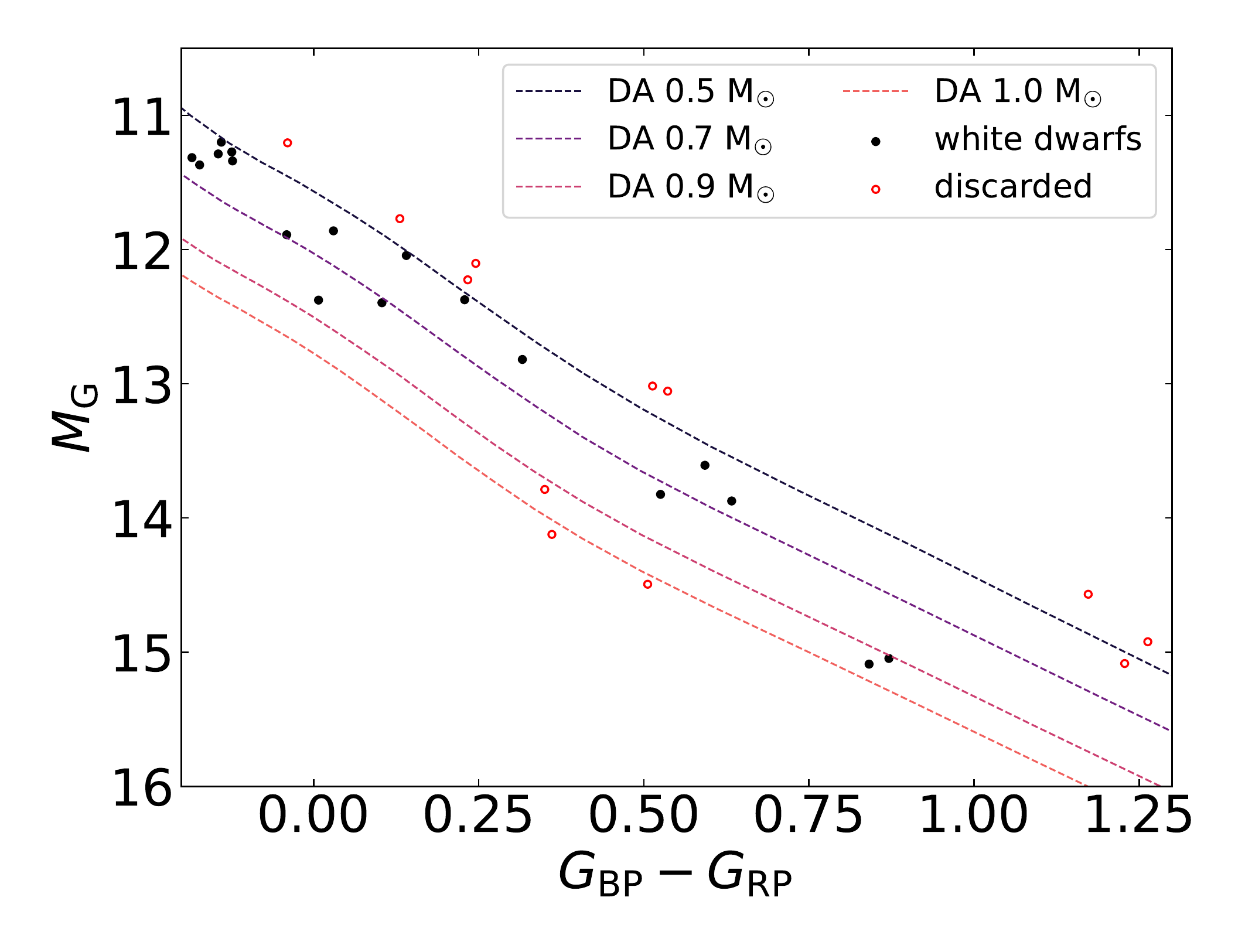}
\caption{Color-magnitude diagram for the $30$ white dwarf which are candidate wide co-movers with M~dwarfs \citep{Kiman2021}. We show as red empty circles the white dwarfs which were discarded because they are outside the ranges which are likely to have followed single star evolution:  $0.5\,\Msun<m_{\rm f}<1.1\,\Msun$, and also white dwarfs with $G_{\rm BP}-G_{\rm RP} < 0.9$ to avoid contamination from non-single white dwarfs. The only object within these limits which is also discarded did not have an estimation of $T_{\rm eff}$ and $\log g$ in \citet{Fusillo2021}. We also show the models of constant mass $0.5\,\Msun$, $0.7\,\Msun$, $0.9\,\Msun$ and $1\,\Msun$ for DA white dwarfs \citep{Bergeron1995,Fontaine2001,Holberg2006,Kowalski2006,Bergeron2011,Tremblay2011,Blouin2018}.} 
\label{fig:cmd_binaries}
\end{center}
\end{figure}

All of the $30$ objects have a white dwarf probability $>95\%$ assigned by \citet{Fusillo2021}.
However, one of the white dwarfs does not have an estimation for $T_{\rm eff}$ and $\log g$.
We discarded that source along with the white dwarfs with masses $< 0.5\,\Msun$, or $> 1.1\,\Msun$, according to their position in the color-magnitude diagram (Figure~\ref{fig:cmd_binaries}) and white dwarfs with $G_{\rm BP}-G_{\rm RP} > 0.9$. 
In short, the reasons for these cuts are that high- and low-mass white dwarfs are thought to be the result of binary evolution, and therefore do not follow the single star evolution assumption made by the cooling tracks included in \texttt{wdwarfdate} \citep{Bedard2020}. 
In addition, the color-cut avoids contamination from non-white dwarfs such as unresolved binaries. 
For a more detailed discussion of the limitations of \texttt{wdwarfdate} see Section~\ref{sec:constraints}. 

We used \texttt{wdwarfdate} to estimate the ages of the remaining $18$ white dwarfs. 
Five of these objects were identified as DA white dwarfs in the MWDD, two as DB, and we assumed DA for the rest, and used the corresponding $T_{\rm eff}$ and $\log g$ from \citet{Fusillo2021}.
Given that most white dwarfs are DAs ($\sim$\,$75\%$), this is a good assumption \citep{Fontaine2001,Fusillo2015}.
To run our code we used the cooling sequence with thick H outermost layer for DA white dwarfs and with thin layer for non-DA, the stellar evolution models for ${\rm [Fe/H]}=0$ and ${\rm v/vcrit}=0.0$, and the \citet{Cummings2018} MIST-based IFMR.
We also calculated these white dwarf ages using the \citet{Marigo2020} IFMR and found similar results, which are not displayed here.
The Gaia source id of the white dwarfs, the input $T_{\rm eff}$ and $\log g$, and the parameters estimated with \texttt{wdwarfdate} are in Table~\ref{table:wd_sample}.
The reported values in this table are the median, $16^{\rm th}$, and $84^{\rm th}$ percentile as uncertainties. 
We were able to estimate the total age for $13$ of the $18$ white dwarfs; the rest have masses outside of the allowed range by the IFMR. 
For these $13$ white dwarfs we used both the Bayesian and fast-test methods included in \texttt{wdwarfdate}, and both results are shown in the table, while for the remaining $5$ we only used the fast-test method to estimate the final mass and cooling age.

The white dwarfs from Table~\ref{table:wd_sample} exemplify the impact of the $T_{\rm eff}$ and $\log g$ uncertainties in the estimation of the parameters, and show the advantage of using the Bayesian method.
For example, Gaia EDR3 $1323951092358466304$ has a $15\%$ and $7\%$ uncertainty on $T_{\rm eff}$ and $\log g$, respectively.
Although the parameters estimated for this white dwarf are within the uncertainties between the fast-test and the Bayesian method, there is a significant difference between, for example, the final masses ($0.64^{+0.16}_{-0.06}\,\Msun$ using the Bayesian method and $0.97^{+0.23}_{-0.26}\,\Msun$ using the fast-test method). 
In this case where the likelihood is not well constrained due to the uncertainties, the prior in the Bayesian method (in particular the IMF) is adding extra information preferring lower-mass white dwarfs, while the fast-test method selects the highest likelihood value.
This was discussed in detailed in Section~\ref{subsec:fasttestmethod}, and will be discussed further in Section~\ref{sec:constraints}.

\begin{deluxetable*}{lccccccccc}[ht!]
\tablewidth{290pt}
\tabletypesize{\scriptsize}
\tablecaption{Candidate white dwarf co-movers with M dwarfs. \label{table:wd_sample}}
\tablehead{                          \colhead{Name}                        & \colhead{SPT\tablenotemark{a}}                        & \colhead{$T_{\rm eff}$ [K]\tablenotemark{b}}                        & \colhead{$\log g$\tablenotemark{b}}                        & \colhead{Met.\tablenotemark{c}}                        & \colhead{$t_{\rm ms}$ [Gyr]}                        & \colhead{$t_{\rm cool}$ [Gyr]}                        & \colhead{$t_{\rm tot}$ [Gyr]}                        & \colhead{$m_{\rm i}$ [M\textsubscript{\(\odot\)}]}                        & \colhead{$m_{\rm f}$ [M\textsubscript{\(\odot\)}]}                        
}\startdata 
Gaia EDR3 2643218398328129664&\nodata&$4661\pm 251$&$7.90\pm 0.23$& FT&\nodata&$5.28_{-1.89}^{+2.79}$&\nodata&\nodata&$0.53_{-0.13}^{+0.14}$\\ 
SDSS J123304.34+030245.6&\nodata&$4735\pm 263$&$7.83\pm 0.27$& FT&\nodata&$4.33_{-1.53}^{+3.13}$&\nodata&\nodata&$0.48_{-0.13}^{+0.16}$\\ 
Gaia EDR3 1323951092358466304&\nodata&$5652\pm 842$&$8.49\pm 0.58$& B&$1.67_{-1.31}^{+4.05}$&$5.42_{-2.35}^{+2.75}$&$8.20_{-2.76}^{+3.32}$&$1.81_{-0.58}^{+1.49}$&$0.64_{-0.06}^{+0.16}$\\ 
& & & & FT &$0.14_{-0.09}^{+0.42}$&$6.09_{-2.00}^{+2.68}$&$6.51_{-1.81}^{+2.72}$&$4.58_{-1.78}^{+2.19}$&$0.96_{-0.25}^{+0.23}$\\ 
Gaia EDR3 1393854635743747200&\nodata&$6469\pm 617$&$8.09\pm 0.32$& B&$2.04_{-1.46}^{+5.28}$&$2.90_{-1.01}^{+2.07}$&$5.96_{-2.07}^{+4.30}$&$1.68_{-0.54}^{+1.11}$&$0.63_{-0.05}^{+0.09}$\\ 
& & & & FT &$0.47_{-0.28}^{+1.70}$&$3.40_{-1.38}^{+1.92}$&$4.56_{-1.24}^{+1.84}$&$2.97_{-1.32}^{+1.13}$&$0.74_{-0.12}^{+0.17}$\\ 
SDSS J155516.95+315307.3&DA&$6740\pm 108$&$8.02\pm 0.05$& B&$4.43_{-2.59}^{+4.86}$&$1.78_{-0.13}^{+0.15}$&$6.17_{-2.47}^{+4.80}$&$1.32_{-0.26}^{+0.42}$&$0.60_{-0.02}^{+0.03}$\\ 
& & & & FT &$3.31_{-1.59}^{+4.09}$&$1.85_{-0.14}^{+0.16}$&$5.11_{-1.38}^{+4.11}$&$1.44_{-0.31}^{+0.35}$&$0.60_{-0.02}^{+0.03}$\\ 
SDSS J114850.50+325406.5&DA&$6981\pm 977$&$7.71\pm 0.57$& FT&\nodata&$1.31_{-0.62}^{+2.24}$&\nodata&\nodata&$0.47_{-0.19}^{+0.32}$\\ 
Gaia EDR3 4467448853280423296&\nodata&$7151\pm 935$&$8.27\pm 0.45$& B&$1.53_{-1.22}^{+5.19}$&$2.68_{-1.16}^{+2.49}$&$5.36_{-2.08}^{+4.41}$&$1.87_{-0.70}^{+1.57}$&$0.65_{-0.06}^{+0.18}$\\ 
& & & & FT &$0.27_{-0.19}^{+0.92}$&$3.41_{-1.71}^{+1.59}$&$3.99_{-1.30}^{+1.65}$&$3.59_{-1.44}^{+2.06}$&$0.86_{-0.19}^{+0.22}$\\ 
Gaia EDR3 860411038927355264&\nodata&$8557\pm 1021$&$8.06\pm 0.41$& B&$2.06_{-1.66}^{+6.18}$&$1.29_{-0.42}^{+1.06}$&$3.90_{-1.60}^{+5.51}$&$1.67_{-0.58}^{+1.48}$&$0.63_{-0.06}^{+0.15}$\\ 
& & & & FT &$0.43_{-0.30}^{+1.72}$&$1.52_{-0.61}^{+1.49}$&$2.60_{-0.85}^{+1.79}$&$3.07_{-1.42}^{+1.61}$&$0.76_{-0.14}^{+0.21}$\\ 
Gaia EDR3 3221072914762838144&\nodata&$9249\pm 492$&$7.72\pm 0.19$& FT&\nodata&$0.56_{-0.13}^{+0.17}$&\nodata&\nodata&$0.45_{-0.09}^{+0.10}$\\ 
SDSS J003000.03-002738.9&DA&$9359\pm 1061$&$7.94\pm 0.45$& FT&$2.40_{-1.96}^{+6.41}$&$0.97_{-0.30}^{+0.67}$&$3.59_{-1.61}^{+5.91}$&$1.59_{-0.51}^{+1.46}$&$0.62_{-0.05}^{+0.14}$\\ 
Gaia EDR3 793351038769083776&\nodata&$10286\pm 1114$&$7.93\pm 0.41$& FT&$2.67_{-2.13}^{+6.33}$&$0.73_{-0.22}^{+0.40}$&$3.45_{-1.74}^{+6.10}$&$1.54_{-0.47}^{+1.31}$&$0.62_{-0.05}^{+0.11}$\\ 
Gaia EDR3 636417842920590208&\nodata&$12338\pm 2743$&$8.45\pm 0.44$& B&$1.13_{-0.97}^{+5.10}$&$1.18_{-0.73}^{+2.65}$&$3.36_{-1.91}^{+5.59}$&$2.19_{-1.00}^{+2.19}$&$0.67_{-0.08}^{+0.27}$\\ 
& & & & FT &$0.17_{-0.10}^{+0.47}$&$0.86_{-0.48}^{+1.27}$&$1.36_{-0.58}^{+1.36}$&$4.30_{-1.62}^{+1.94}$&$0.93_{-0.23}^{+0.21}$\\ 
Gaia EDR3 2536705752006690304&\nodata&$12371\pm 2799$&$8.07\pm 0.50$& B&$2.05_{-1.69}^{+5.70}$&$1.00_{-0.61}^{+2.71}$&$4.40_{-2.68}^{+5.81}$&$1.68_{-0.56}^{+1.61}$&$0.63_{-0.06}^{+0.17}$\\ 
& & & & FT &$0.35_{-0.25}^{+1.20}$&$0.64_{-0.35}^{+0.97}$&$1.32_{-0.54}^{+1.61}$&$3.29_{-1.44}^{+1.97}$&$0.80_{-0.16}^{+0.23}$\\ 
SDSS J082233.92+213047.3&DA&$12859\pm 1751$&$8.18\pm 0.22$& B&$1.82_{-1.37}^{+5.43}$&$0.46_{-0.17}^{+0.32}$&$2.38_{-1.24}^{+5.51}$&$1.75_{-0.61}^{+1.27}$&$0.64_{-0.06}^{+0.11}$\\ 
& & & & FT &$0.47_{-0.22}^{+1.20}$&$0.46_{-0.17}^{+0.28}$&$1.09_{-0.32}^{+1.06}$&$2.97_{-1.17}^{+0.77}$&$0.74_{-0.11}^{+0.13}$\\ 
Gaia EDR3 703753485491279488&\nodata&$14390\pm 2095$&$7.83\pm 0.28$& FT&\nodata&$0.18_{-0.09}^{+0.16}$&\nodata&\nodata&$0.51_{-0.12}^{+0.15}$\\ 
RMB 103&DB&$14096\pm 518$&$7.93\pm 0.08$& B&$6.23_{-3.61}^{+4.83}$&$0.25_{-0.03}^{+0.04}$&$6.50_{-3.62}^{+4.77}$&$1.19_{-0.18}^{+0.36}$&$0.58_{-0.03}^{+0.03}$\\ 
& & & & FT &$5.38_{-3.28}^{+5.40}$&$0.25_{-0.03}^{+0.04}$&$5.62_{-3.23}^{+5.39}$&$1.24_{-0.23}^{+0.42}$&$0.59_{-0.02}^{+0.03}$\\ 
LAMOST J095620.88+272729.8&DA&$15137\pm 682$&$7.97\pm 0.08$& B&$4.52_{-2.82}^{+5.23}$&$0.20_{-0.03}^{+0.04}$&$4.72_{-2.81}^{+5.21}$&$1.31_{-0.26}^{+0.48}$&$0.60_{-0.03}^{+0.03}$\\ 
& & & & FT &$2.91_{-1.54}^{+4.82}$&$0.20_{-0.03}^{+0.04}$&$3.10_{-1.51}^{+4.79}$&$1.50_{-0.38}^{+0.50}$&$0.61_{-0.03}^{+0.04}$\\ 
PB  6042&DB&$15602\pm 2052$&$8.11\pm 0.30$& B&$2.49_{-1.89}^{+6.10}$&$0.26_{-0.10}^{+0.19}$&$2.80_{-1.79}^{+6.02}$&$1.57_{-0.49}^{+1.16}$&$0.62_{-0.05}^{+0.10}$\\ 
& & & & FT &$0.49_{-0.26}^{+1.74}$&$0.29_{-0.12}^{+0.21}$&$0.89_{-0.26}^{+1.57}$&$2.92_{-1.29}^{+0.94}$&$0.73_{-0.11}^{+0.15}$\\ 
\enddata 
\tablenotetext{a}{Classification in the Montreal White Dwarf Data Base.}
\tablenotetext{b}{Parameters from \citet{Fusillo2019}, estimated from photometry.}
\tablenotetext{c}{Method used in \texttt{wdwarfdate}: B = Bayesian, F = Fast-test.}
\end{deluxetable*}

\section{Limitations and assumptions}
\label{sec:constraints}

White dwarf cosmochronology is a powerful tool to estimate stellar ages. However, it is based on a number of models and assumptions that need to be taken into account. 
In this section, we discuss the constraints and model choices when using \texttt{wdwarfdate}.

\begin{figure}[ht!]
\begin{center}
\includegraphics[width=\linewidth]{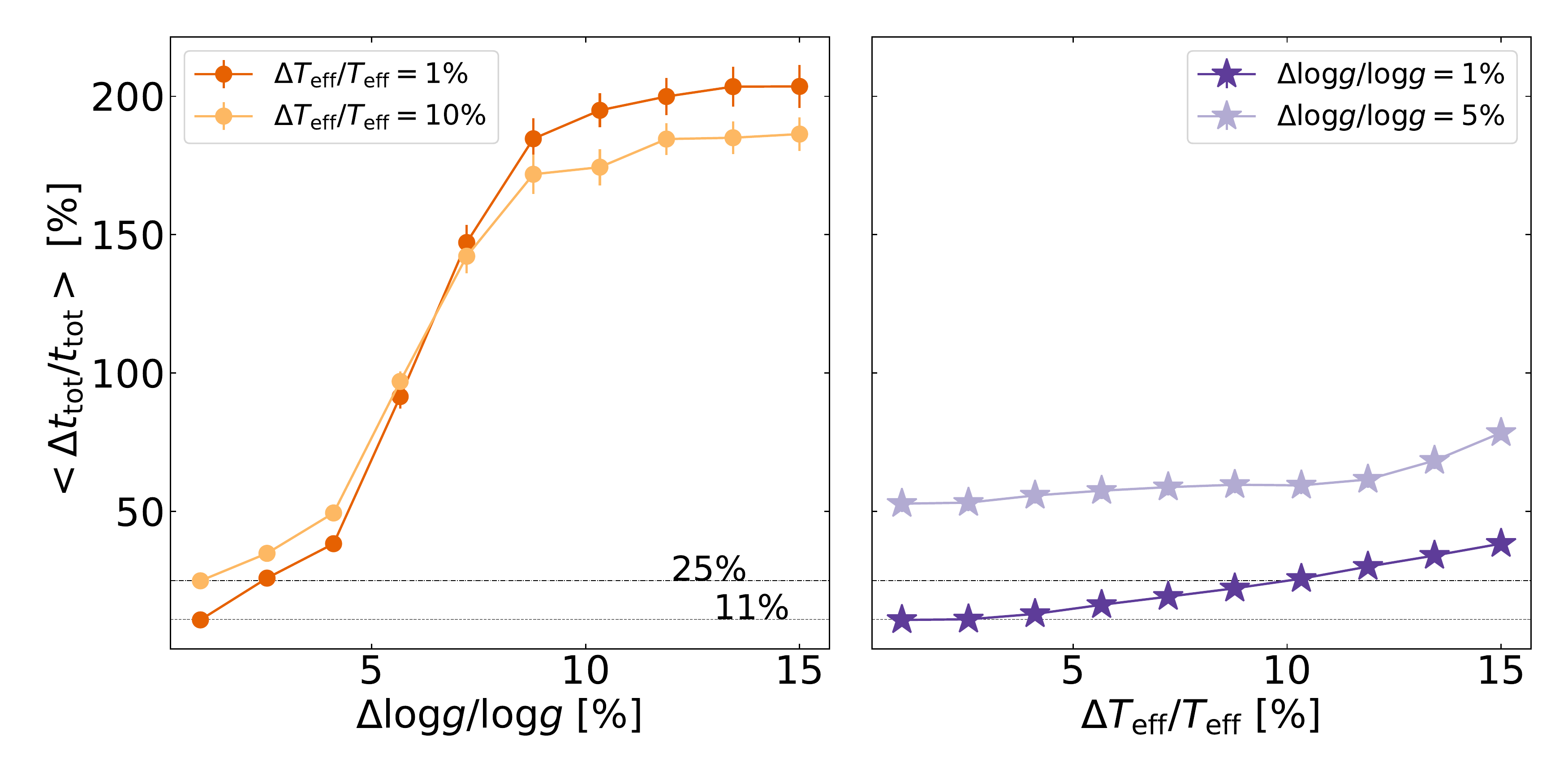}
\caption{Median relative uncertainty in the total age calculated using the Bayesian method in \texttt{wdwarfdate}. The median is calculated over all $T_{\rm eff}$ and $\log g$ results with the same input uncertainties, and the uncertainty in the median is calculated as the standard deviation. In the left panel, the relative uncertainty in $T_{\rm eff}$ is fixed at $1\%$ and $10\%$. In the right panel, the relative uncertainty in $\log g$ is fixed at $1\%$ and $5\%$.} 
\label{fig:relative_error}
\end{center}
\end{figure}

We run a grid of $100$ white dwarfs with $1,500\,{\rm K} <T_{\rm eff}<100,000$\,K and $7<\log g<9.3$ with different relative uncertainties to test the resulting uncertainty of the Bayesian total ages derived with \texttt{wdwarfdate}. In this study we used the \citet{Cummings2018} MIST-based IFMR, with the DA white dwarf cooling sequences and ${\rm [Fe/H]} = 0$ and ${\rm v/vcrit}=0$ stellar evolution models. We found that for uncertainties of $1\%$ in both $T_{\rm eff}$ and $\log g$, the median relative uncertainty in total age is $\sim 11\%$, as shown in Figure~\ref{fig:relative_error}. If the uncertainty of $T_{\rm eff}$ is increased to $10\%$, the total age uncertainty increases to $\sim 25\%$. On the other hand, using an uncertainty of $1\%$ for $T_{\rm eff}$ and $5\%$ for $\log g$, the total age uncertainty increases to $\sim 52\%$, showing the importance of precise input measurements, in particular of $\log g$. For a study of the total age relative uncertainty as a function of $T_{\rm eff}$ and $\log g$ see Appendix~\ref{sec:relative_unc}. We repeated this analysis for non-DA cooling models and found no significant difference in the results.

\begin{figure}[ht!]
\begin{center}
\includegraphics[width=\linewidth]{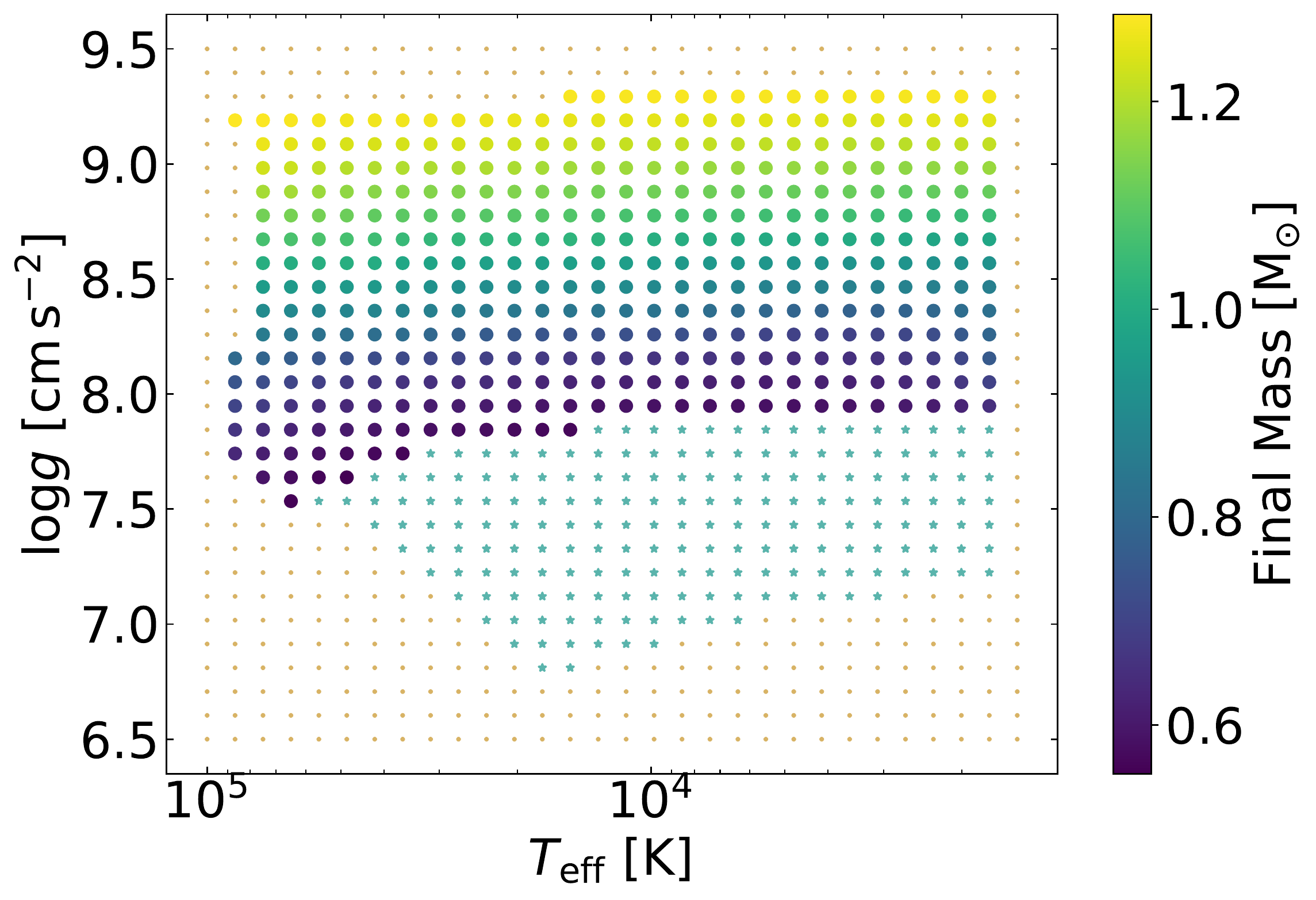}
\caption{Limits of $T_{\rm eff}$ and $\log g$ for \texttt{wdwarfdate}: In brown are the values for which the code does not estimate any parameters because they are outside of the allowed range by the cooling tracks. In green-blue five-point stars are the values for which only a final mass and cooling age are estimated using the fast-test method, but not the rest of the parameters because of the limits where the MIST-based \citet{Cummings2018} IFMR (the chosen relation for this example) is calibrated. We color-coded by final mass the values for which \texttt{wdwarfdate} estimates all the parameters using the Bayesian method: final mass, cooling age, initial mass, main sequence age and total age.} 
\label{fig:limits_wdwarfdate}
\end{center}
\end{figure}

In addition to the limits on $T_{\rm eff}$ and $\log g$ set by the cooling tracks discussed in Section~\ref{subsec:modelsincluded}, the IFMR imposes extra limits to the input values for which \texttt{wdwarfdate} will be able to estimate a total age, main sequence age and initial mass. 
To quantify this limit, we run \texttt{wdwarfdate} on a grid of $T_{\rm eff}$ and $\log g$ and recorded which parameters were estimated.
In this exercise, we adopted the MIST-based \citet{Cummings2018} IFMR, the cooling sequences for DA white dwarfs, and the stellar evolution models for ${\rm [Fe/H]}=0$ and ${\rm v/vcrit} = 0.0$.  
The values of $T_{\rm eff}$ and $\log g$ for which \texttt{wdwarfdate} can estimate a total age are color-coded by final mass in Figure~\ref{fig:limits_wdwarfdate}. 
In summary, \texttt{wdwarfdate} will derive Bayesian total ages for $1,500\,{\rm K}  \lesssim T_{\rm eff}  \lesssim 90,000$\,K and $7.9  \lesssim \log g  \lesssim 9.3$, and only final mass and cooling age using the fast-test method for $7.0 \lesssim \log g  \lesssim 7.9$ (green-blue five point stars in Figure~\ref{fig:limits_wdwarfdate}).

The masses for C/O core white dwarfs ranges from $0.45-1.1\,\Msun$. For lower masses, the core is composed of He and for higher masses of O and/or Ne \citep{Fontaine2001,Garcia-berro1997}. 
As we implemented cooling tracks that assume a  C/O core in \texttt{wdwarfdate}, parameter estimation using our code outside this range should be taken with caution. 
We discuss further the lower and upper mass limits below.

Within a Hubble time, it is not possible to have a white dwarf with a mass smaller than $0.45\,\Msun$ that was formed by single stellar evolution. 
Low-mass white dwarfs ($<0.45\,\Msun$) cannot ignite He, therefore they have a He core, and are thought to be the result of binary evolution \citep[e.g.,][]{Marsh1995,Brown2011}. 
There are studies which propose that these objects could also be formed from a red giant branch star which had a high mass loss rate \citep[e.g., ][]{Kilic2007}.
In addition, binary stars could also appear to be white dwarfs of masses $<0.45\,\Msun$, for example unresolved white dwarf binaries such as double degenerate stars, look like one single over-luminous star. 
With the same $T_{\rm eff}$, this means bigger radius, which means smaller mass according to the mass-radius relation of degenerate objects \citep{Bergeron2001,Bedard2017,Kilic2020}. 
This phenomenon could mimic a white dwarf of mass $<0.45\,\Msun$.

Higher-mass white dwarfs ($>1.1\,\Msun$) should also be viewed with caution. 
These stars are potentially magnetic white dwarfs and may have come from a merger \citep{Ferrario2015}, which would make the single star evolution assumption fail, and therefore the models would not reproduce the true evolution of the object.

The typical structure of a white dwarf consists of a C/O core with a thin shell of He, surrounded by a thin layer of H. 
Both the core and the surface layers have an important role in how the white dwarfs cool: the thermal energy is stored in the core while the heat outflow is regulated by the opaque envelope \citep{Hansen2004}. 
For example, DA white dwarfs with thick H layers cool slower than non-DA objects because the H envelope is a better insulator \citep{Fontaine2001}. 
In addition, the chemical composition of the core will affect the heat capacity of the object, and therefore the cooling rate \citep{Hansen2004}.
For instance, the estimated cooling age of a white dwarf assuming a pure C core is an upper limit, while assuming a pure O core is a lower limit, with differences of a few Gyr for the oldest objects \citep{Fontaine2001}. 
Although the core composition is still a source of uncertainty for white dwarf cosmochronology \citep{Simon2015}, it is necessary to assume one to estimate the cooling age and mass of a white dwarf with evolutionary models. \texttt{wdwardate} uses cooling tracks from \citet{Bedard2020} which assume a C/O core with the same proportions of C and O with the option of choosing between a thin or thick outermost H layer.

The parameter estimation using \texttt{wdwarfdate} depends also on the relation between the progenitor mass and the white dwarf mass, meaning the IFMR. 
This is an empirical relation, which is improving as more precise measurements of the parameters of white dwarfs and their progenitors become available. 
The IFMRs included in \texttt{wdwarfdate} are similar to each other (See Figure~\ref{fig:IFMRcomparison}), and we have seen no significant difference in the median values of the ages and masses estimated in this study when changing the IFMR. 
However, the IFMR should be chosen carefully when using \texttt{wdwarfdate} because the shape of the posterior PDF can be affected by it.
Our recommendation among the available IFMRs is to use the MIST-based \citet{Cummings2018} relation, because this is the most complete and simplest relation to use with our method. 
Although \citet{Marigo2020} is more recent than \citet{Cummings2018}, the difference is small between the relations, and the kink in the former complicates the sampling of the posterior without significant difference in the median mass or age of the white dwarf. 

The main sequence age of the progenitor is obtained from a stellar evolution model, based on its mass, abundances and rotation, which can modify the estimation of the total age \citep{2022arXiv220308971M}. 
We tested the parameter estimation for $50$ randomly selected white dwarfs with $1500\,{\rm K}<T_{\rm eff}<100,000\,{\rm K}$ and $7<\log g<9.3$, with uncertainties of $10\%$ and $1\%$ respectively, using all the possible combinations of ${\rm [Fe/H]=\{-4,-1,0.5\}}$ (dex) and ${\rm v/vcrit = \{0.0,0.4\}}$.
For all the runs we used the MIST-based \citet{Cummings2018} IFMR, and the cooling sequences for DA white dwarfs.
We compared the results to the parameters estimated using ${\rm [Fe/H]}=0$, and ${\rm v/vcrit} = 0.0$ to estimate the effect of using different stellar evolution models.
We found no significant change in the estimated values of the final mass or cooling age, which is expected because we did not modify the parameters of the white dwarf. 
For the initial mass we found a small difference of up to $0.16\,\Msun$ when ${\rm [Fe/H]}=-4$ for both cases of ${\rm v/vcrit}$.
For the main sequence age and total age we found differences of up to $0.7$\,Gyr when ${\rm [Fe/H]}=-4$, again not changing significantly with ${\rm v/vcrit}$.
This shows that the parameters of the progenitor star need to be chosen carefully given that they can affect the results.
In case there is no information of the parameters of the progenitor star, it is reasonable to choose solar metallicity (${\rm [Fe/H]}=0$), taking into account the possible variations described above. 

The last aspect to take into account when running \texttt{wdwarfdate} is the choice of priors. Currently the code assumes a constant SFH and the IMF to be $m_{\rm i}^{-2.3}$ when using the Bayesian mode (see Section~\ref{subsec:methods}), and it does not allow changes in these priors. As discussed in Sections~\ref{subsec:fasttestmethod} and \ref{subsec:calc_ages}, when the uncertainty in the input parameters is large, the likelihood will be less constrained and the priors will have a bigger effect, making the initial and the final masses smaller, and therefore affecting the estimated ages as well. In this case, the user could consider running the fast-test method instead of the Bayesian method. In this alternative method, \texttt{wdwarfdate} will perform a Monte Carlo uncertainty propagation to estimate the white dwarf parameters (as it was described in Section~\ref{subsec:fasttestmethod}) and will not be affected by the priors.



\section{Summary}
\label{sec:summary}

In this work we presented \texttt{wdwarfdate}, a publicly available Python package which derives Bayesian total ages of white dwarfs from an effective temperature ($T_{\rm eff}$) and a surface gravity ($\log g$), assuming single star evolution. 
\texttt{wdwarfdate} obtains the probability distributions by explicitly integrating over the likelihood and priors to perform the normalization for the following parameters: mass of the progenitor star, cooling age of the white dwarf and $\Delta _{\rm m}$, which models the scatter in the IFMR. 
From these parameters, the code obtains probability distributions for the mass of the white dwarf, the main sequence age of the progenitor and the total age.
The derived ages depend on the input parameters and on the models chosen when initializing \texttt{wdwarfdate}, which are summarize in Table~\ref{table:sum_models} (See Appendix~\ref{sec:run_wdwarfdate} for an example on how to use \texttt{wdwarfdate}).

We tested \texttt{wdwarfdate} on white dwarfs cluster members from \citet{Cummings2018} and \citet{Canton2021}.
In general we found good agreement between our results and theirs for the mass and age of the white dwarf as well as the properties of the progenitor star. 
In addition, we used \texttt{wdwarfdate} to estimate the ages of a set of $18$ white dwarfs which are candidate co-movers with M~dwarfs \citep{Kiman2021}, and therefore conform a group of low-mass star age-calibrators.

We also described a detailed analysis of the constraints on the code which need to be taken into account before using. 
In summary: 
\begin{itemize}
    \item We found a typical uncertainty of $10\%$ on the Bayesian total age for $T_{\rm eff}$ and $\log g$ with uncertainties of $1\%$, and of $25\%$ with input uncertainties of $10\%$ and $1\%$ for $T_{\rm eff}$ and $\log g$, respectively.
    \item Combining the limitation of the cooling tracks with the restrictions of the IFMR, the values for which \texttt{wdwarfdate} can derive Bayesian total ages are $1,500\,{\rm K}  \lesssim T_{\rm eff}  \lesssim 90,000$\,K and $7.9  \lesssim \log g  \lesssim 9.3$ (cm ${\rm s^{-2}}$).
    \item Given that the cooling tracks assume single star evolution and C/O core for the white dwarfs, the best range of final masses to use \texttt{wdwarfdate} is $0.45-1.1\,\Msun$, because objects outside this range are not likely to have evolved as a single star.
    \item The IFMRs included in our code are similar to each other and the results are not modified significantly from using one over the other. However we recommend using the MIST-based \citet{Cummings2018} IFMR, given it is one of the most complete studies of white dwarfs in clusters currently available, and has a relatively simple functional form.
    \item The choice of ${\rm [Fe/H]}$ can modify the main sequence age and total age by up to $0.7$\,Gyr, but if there is no information about the progenitor star available, solar metallicity (${\rm [Fe/H]}=0$) is a reasonable assumption.
    \item When the uncertainty of the input parameters is large, the prior in the Bayesian method will affect the estimation of the age given that the likelihood is not well constrained. The user can consider using the fast-test method in these cases.
\end{itemize}

Although Gaia has increased the number of known white dwarfs by a factor of $10$, we still need more white dwarfs which belong to known clusters with spectra to improve the calibration of the IFMR. 
This relation is one of the main limitations when estimating the total age of a white dwarf because the complete functional form of the IFMR is still not known with great accuracy, especially in the low-mass regime.

In future work we plan to include the estimation of $T_{\rm eff}$ and $\log g$ from photometry in \texttt{wdwarfdate}, and extra priors in case the final mass or total age of the white dwarf are known, to improve the estimation of the other parameters. 

To learn about how to use \texttt{wdwarfdate}, check out the source code \citep{rocio_kiman_2022_6633759}\footnote{\url{https://github.com/rkiman/wdwarfdate.git}} and the online documentation\footnote{\url{https://wdwarfdate.readthedocs.io}}.

\end{CJK}

\section{Acknowledgements}
The authors would like to thank Jeff Andrews, Lars Bildsten, Javier Roulet and Antoine B\'edard for useful discussions.

Support for this project was provided by a PSC-CUNY Award, jointly funded by The Professional Staff Congress and The City University of New York.

This material is based upon work supported by the National Science Foundation under Grant No. 1614527.

This work has been supported by NASA K2 Guest Observer program under award 80NSSC19K0106.

This work was supported by the SDSS Faculty and Student Team (FAST) initiative.

Support for this work was provided by the William E Macaulay Honors College of The City University of New York

S.X. is supported by the international Gemini Observatory, a program of NSF's NOIRLab, which is managed by the Association of Universities for Research in Astronomy (AURA) under a cooperative agreement with the National Science Foundation, on behalf of the Gemini partnership of Argentina, Brazil, Canada, Chile, the Republic of Korea, and the United States of America.

J.F. and S.X. acknowledge support from the Heising-Simons Foundation.

S.L.C. acknowledges the support of an STFC Ernest Rutherford Research Fellowship.

\software{\texttt{scipy} \citep{2020SciPy-NMeth}; \texttt{numpy} \citep{oliphant2006guide,van2011numpy}; \texttt{matplotlib} \citep{Hunter2007}; \texttt{astropy} \citep{astropy2013,astropy2018}} 

\appendix

\section{Running \texttt{wdwarfdate}}
\label{sec:run_wdwarfdate}

Below we include an example of how to run \texttt{wdwarfdate}, where we selected a white dwarf from \citet{Cummings2018}, and we are using the cooling tracks for a DA white dwarf, the stellar evolution track corresponding to [Fe/H] = 0 and no rotation, and the MIST-based IFMR from \citet{Cummings2018}. 
By default \texttt{wdwarfdate} will run the Bayesian method, but the fast-test method can be selected by indicating \texttt{method='fast\_test'} in the \texttt{WhiteDwarf} class.
The object \texttt{results} is an \texttt{astropy} Table\footnote{\url{https://docs.astropy.org/en/stable/table/index.html}} with the total age of the object, main sequence age and mass of the progenitor and cooling age and mass of the white dwarf, and their uncertainties. 
This example takes $21$ seconds to run in a personal computer.
For more information on how to run \texttt{wdwarfdate} see the documentation of the code.

\begin{verbatim}

import wdwarfdate

teff = 19250
teff_err = 500
logg = 8.16
logg_err = 0.084

WD = wdwarfdate.WhiteDwarf(teff,teff_err,logg,logg_err,
                           model_wd='DA',
                           feh='p0.00',
                           vvcrit='0.0',
                           model_ifmr = 'Cummings_2018_MIST')
WD.calc_wd_age()

results = WD.results

\end{verbatim}

\section{Relative uncertainty in total age.}
\label{sec:relative_unc}

\begin{figure}[ht!]
\begin{center}
\includegraphics[width=\linewidth]{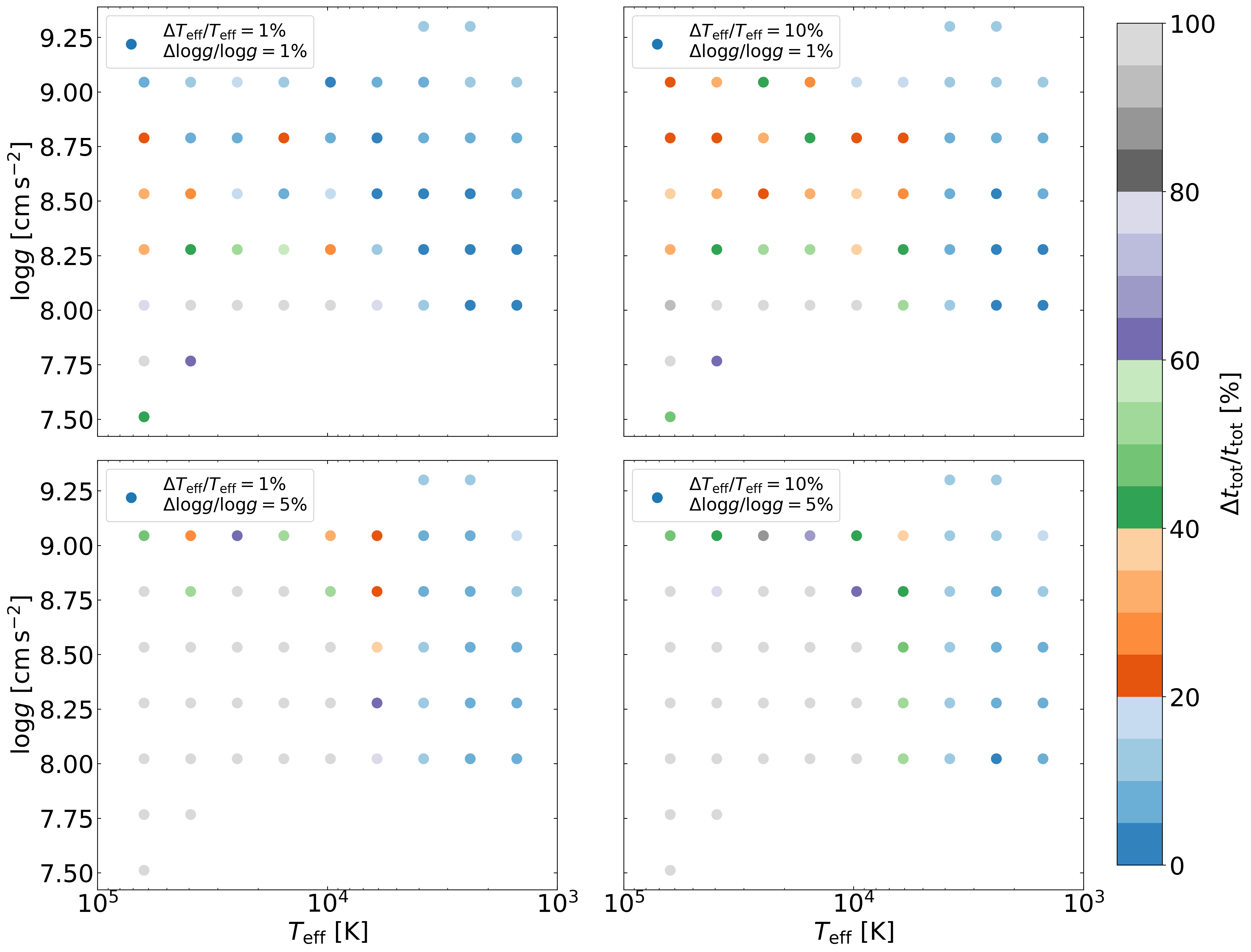}
\caption{Relative uncertainty in the Bayesian total age estimated using \texttt{wdwarfdate}. We include four cases of different input uncertainties to show the variation as a function of $T_{\rm eff}$ and $\log g$.} 
\label{fig:relative_error_teff_logg}
\end{center}
\end{figure}

To study the relative uncertainty in the Bayesian total age estimated with \texttt{wdwarfdate} we run a grid of $100$ white dwarfs with $1,500\,{\rm K} <T_{\rm eff}<100,000$\,K and $7<\log g<9.3$, with different uncertainties. 
In this study we used the \citet{Cummings2018} MIST-based IFMR, with the DA white dwarf cooling sequences and ${\rm [Fe/H]} = 0$ and ${\rm v/vcrit}=0$ stellar evolution models.
In Figure~\ref{fig:relative_error_teff_logg} we show the relative uncertainty in the total age ($t_{\rm tot}$) for four cases of input uncertainties: 
$\Delta T_{\rm eff}/T_{\rm eff}=1\%$ and $\Delta \log g/\log g=1\%$, 
$\Delta T_{\rm eff}/T_{\rm eff}=10\%$ and $\Delta \log g/\log g=1\%$, 
$\Delta T_{\rm eff}/T_{\rm eff}=1\%$ and $\Delta \log g/\log g=5\%$, and 
$\Delta T_{\rm eff}/T_{\rm eff}=10\%$ and $\Delta \log g/\log g=5\%$.
As discussed in Section~\ref{sec:constraints}, the biggest effect on the uncertainty of $t_{\rm tot}$ comes from increasing the uncertainty of $\log g$. In addition, we show that higher $T_{\rm eff}$ are the most affected by the increase in the uncertainty of the input parameters.
We repeated this analysis for non-DA cooling models and found no significant difference.

\bibliographystyle{aasjournal}
\bibliography{references,references_extra}

\end{document}